\documentclass[fleqn,usenatbib,useAMS]{mnras}
\usepackage{pifont}
\usepackage{amsmath}
\usepackage{graphicx,epstopdf}
\DeclareGraphicsExtensions{.ps}
\usepackage[utf8]{inputenc}
\usepackage{url}
\usepackage{lscape}
\usepackage{float}
\usepackage[section]{placeins}
\usepackage{color}

\usepackage{newtxtext,newtxmath}

\usepackage[T1]{fontenc}
\usepackage{ae,aecompl}

\usepackage{amssymb}	
\usepackage{longtable}
\usepackage{pdflscape}
\usepackage{ulem}

\newcommand{\teff}{$T_\textrm{eff}$}
\newcommand{\logg}{$\log g$}

\newcommand{\mh}{[M/H]}
\newcommand{\luminosity}{$\log ({L_\star/L_{\odot}})$}
\newcommand{\radius}{$R_\star$}
\newcommand{\mass}{$M_\star$}
\newcommand{\kms}{km\,s$^{-1}$}
\newcommand{\vsini}{$\nu$\,sin\,$i$}
\newcommand{\vrad}{$v_{\rm rad}$}
\newcommand{\halpha}{H$\alpha$}
\newcommand{\hbeta}{H$\beta$}

\newcommand{\tess}{{\it TESS}}
\newcommand{\target}{HD\,180347}

\def\spose#1{\hbox to 0pt{#1\hss}}
\def\lta{\mathrel{\spose{\lower 3pt\hbox{$\mathchar"218$}}
     \raise 2.0pt\hbox{$\mathchar"13C$}}}
\def\gta{\mathrel{\spose{\lower 3pt\hbox{$\mathchar"218$}}
     \raise 2.0pt\hbox{$\mathchar"13E$}}}

\title[Characterisation of \target]{High-precision photometric and high-resolution spectroscopic characterisation of \target}

\author[O. Trust et al.]{
Otto Trust$^{1}$\thanks{E-mail: otrust@must.ac.ug},
Lyudmila Mashonkina$^{2}$,
Edward Jurua$^{1}$, 
Peter De Cat$^{3}$,
Vadim Tsymbal$^{2}$,\newauthor
 and Santosh Joshi$^{4}$
 \\
 $^{1}$Department of Physics, Mbarara University of Science and Technology, P.O. Box 1410, Mbarara, Uganda\\
 $^{2}$Institute of Astronomy, Russian Academy of Sciences, 119017, Pyatnitskaya str., 48, Moscow, Russia\\
 $^{3}$Royal Observatory of Belgium, Ringlaan 3, B-1180 Brussel, Belgium\\
 $^{4}$Aryabhatta Research Institute of Observational Sciences, Manora Peak, Nainital- 263002, India
 }

\date{Accepted 2023 June 23. Received 2023 May 09; in original form 2023 February 02}

\pubyear{2023}

\begin{document}
\label{firstpage}
\pagerange{\pageref{firstpage}--\pageref{lastpage}}
\maketitle

\begin{abstract}We report the analysis of high-precision space-based photometric and high-resolution spectroscopic observations of \target. The high-quality light curves from the Transiting Exoplanet Survey Satellite (\tess) under sectors 14, 15, and 26 were used. By visual inspection of the light curves and the Fourier transforms, only low-frequency signals (less than 1\,d$^{-1}$) were detected.
After using wavelet, autocorrelation, and composite spectrum analyses, \target\ is classified as a rotational variable with a period of about 4.1\,$\pm$\,0.2\,days. In reference to the observation limit of \tess, no pulsations were detected. For the spectroscopic analysis, we used data collected with the High Efficiency and Resolution Mercator \'{E}chelle Spectrograph (HERMES). We determined the spectral type of this star and obtained atmospheric parameters such as the effective temperature, the surface gravity, and the projected rotational, microturbulent, and radial velocities. We performed a detailed chemical abundance analysis. The LTE abundances were derived for 25 chemical elements. For 13 of them, including Ca, Sc, Sr, Zr, and Ba, which are important for the characterisation of chemical peculiarity, we also present the non-local thermodynamic equilibrium (NLTE) abundances. NLTE improves the accuracy of the derived abundances and confirms that Ca and Sc are depleted in HD\,180347 relative to their solar abundances, while the heavy elements beyond Sr are enhanced, by more than 0.7\,dex. Based on the spectral class and the element abundance pattern, we classify this star as Am (kA1hA8mA8).
\end{abstract}


Over 10\% of the intermediate mass main-sequence stars (A- and F-type) are chemically peculiar (CP). The CP stars show anomalies in their chemical composition compared to the solar one. The CP stars are categorised into four major groups: CP1 stars (the metallic-line or Am/Fm stars), CP2 stars (the magnetic Ap stars), CP3 stars (the Mercury-Manganese or HgMn stars) and CP4 stars (the He-weak stars), based on their magnetic field and absorption line strengths \citep{1974ARA&A..12..257P}. The Am stars are distinguished by low abundances of some elements such as Ca and Sc, as well as an excess of Fe-group metals and often weak or absent magnetic fields \citep{1970PASP...82..781C, 1974ARA&A..12..257P, 2007AstBu..62...62R}. In the Am stars subgroup, the Ca\,\textsc{ii} K-line appears too early
compared to the types derived from hydrogen lines, whilst metallic lines appear too late, resulting in spectral types inferred from the Ca\,\textsc{ii} K- and metal lines differing by five or more spectral subclasses. For the marginal Am stars, the spectral subclasses between the Ca\,\textsc{ii} K- and metal lines are less than five. The frequently used detailed classification for this class of objects involves three spectral subtypes prefixed with k, h, and m, which represent Ca\,\textsc{ii} K-line, hydrogen lines, and metallic lines, respectively.

Some metals like Si, Cr, Sr, and Hg, as well as rare-earth elements like Eu, Nd, Pr, and others, are overabundant in the CP2 stars in comparison to solar values  \citep{1970PASP...82..781C, 1974ARA&A..12..257P, 2000BaltA...9..253K, 2007AstBu..62...62R}. Unlike CP1 and CP3, the presence of well-organized magnetic fields with strengths of several tens of kG is also a feature of CP2 stars \citep{2007A&A...475.1053A}. In their spectra, the CP3 stars have amplified Hg\textsc{ii} (398.4 nm) and/or Mn\textsc{ii} lines, as well as faint lines of light elements (such as He, Al, and N). In the spectra of the CP4 stars, there are unusually weak He\textsc{i} lines and these stars are characterised by magnetic
fields with a strength of up to 1 kG \citep{1974PASP...86...67J,1974ARA&A..12..257P, 1997A&A...319..928S, 1998ApJ...504..533B}. The CP stars are slow rotators with a projected rotational velocity (\vsini) below 120\,\kms \citep{Abt2009}.

The interplay between radiative levitation and gravitational settling, known as atomic diffusion, is assumed to be the principal source of chemical anomalies in CP stars \citep{1970ApJ...160..641M,1970ApJ...162L..45W,khokh, hui,2000ApJ...529..338R, 2003ASPC..305..199T, 2011sf2a.conf..253T}. In the absence of mixing, light elements sink under gravity and are perceived as under-abundant, whereas heavy elements are radiatively forced outward and reflected as over-abundant. This idea necessitates calm and stable atmospheres, which are aided by CP stars' sluggish rotation behaviour \citep{2008JKAS...41...83T,2008A&A...483..891F, stateva, Abt2009} and Ap stars' strong magnetic fields. Strong magnetic fields are expected to stabilize convective material, but slow rotation reduces meridional circulation and, as a result, reduces mixing, which would prevent atomic diffusion.
The physical processes that are active in CP stars, such as pulsation \citep{1988AcA....38...61D,1998A&A...334..911S} and convective overshooting \citep{2004ApJ...601..512B, 2019MNRAS.485.4641C}, and their chemical abundances \citep{2001astro.ph.11179T, 2014PhDT.......131M} are heavily influenced by rotation.

\citet{2009A&A...498..961R}'s General Catalogue of Ap and Am stars has 8205 peculiar (or suspected peculiar) stars, making it one of the most comprehensive catalogues of peculiar stars. There are 3652 (candidate) Ap stars, 162 (candidate) HgMn stars, 4299 (candidate) Am stars, and 92 stars that have been incorrectly catalogued as Ap, HgMn, or Am at least once. \target\ (= TIC\,298969563 = KIC\,12253106) is one of the probable Am stars in the General Catalogue of Ap and Am stars by \citet{2009A&A...498..961R}.

\citet{smalley} found \target\ to be variable with amplitude less than 0.01\,mmag. Later on, \citet{balona15} classified the variability as rotational with a period of 4.1\,days using data of the nominal {\it Kepler} mission \citep{borucki10}.
Recently, \citet{2019MNRAS.484.2530C} reported \target\ among possible pulsators. However, based on the diffusion theory, the He\,\textsc{ii} ionisation zone, which excites $\delta$ Scuti-type pulsations, should be absent in Am stars. Therefore, it is important to perform a detailed and homogeneous classification and variability study of such a star.

To search and study the pulsational variabilities in Ap and Am/Fm stars, a dedicated ground-based project the ``Nainital-Cape Survey'' was initiated between astronomers of India and South Africa. However, with time, astronomers from other institutions in other countries joined this programme, transforming it into a multi-national collaborative project and a number of results are published \citep[e.g;][]{2000BASI...28..251A, 2001AA...371.1048M,2003MNRAS.344..431J,2006A&A...455..303J, 2009A&A...507.1763J, 2010MNRAS.401.1299J, 2012MNRAS.424.2002J, 2016A&A...590A.116J, 2017MNRAS.467..633J, 2020MNRAS.492.3143T, 2021MNRAS.504.5528T, 2022MNRAS.510.5854J}. In this paper, we study \target\ using data in the \tess\ archive supplemented with high-resolution spectroscopic data from HERMES and spectrophotometric observations available to the public in various databases.  The aim of the study is to investigate the photometric variability and fully characterise the candidate Am star \target.

This paper is organised as follows.  The spectroscopic observations and data reduction are discussed in Section\,\ref{obs}. \tess\ photometry is given in Section\,\ref{tess}. Spectral classification is presented in Section\,\ref{class}, while fundamental parameters and individual chemical abundance analyses are given in Section\,\ref{sect:fund_par} and \ref{sect:abun}, respectively. Finally, the conclusions are given in Section\,\ref{concl}.

\section{Observations and Data Reduction}
\label{obs}

A high-resolution spectroscopic observation of \target\ was done on the night of 6 November 2018, using HERMES \citep{2011A&A...526A..69R} mounted at the Cassegrain focus of the \mbox{1.2-m} Mercator telescope located at La Palma, Spain. This star was observed for a total exposure time of 13.3\,minutes. This spectrograph records optical spectra in the wavelength ($\lambda$) range of 377 to 900\,nm spanning 55 spectral orders in a single exposure. This instrument has a resolving power of 85\,000 in high-resolution mode, with a radial velocity stability of roughly 50\,m\,s$^{-1}$ and an outstanding throughput \citep{2011A&A...526A..69R}.

The spectrum was reduced using the dedicated HERMES pipeline following the usual reduction procedure for \'{e}chelle spectra, including subtraction of bias and stray light, flat-field correction, order-by-order extraction, wavelength calibration frames with Thorium-Argon lamps, removal of cosmic rays, and merging of the orders. This procedure resulted in a spectrum with a signal-to-noise ratio (SNR) of 108, 119, and 76 at $\lambda=500,\,650$, and 810\,nm, respectively. The spectrum was manually normalized up to the local continuum using an integrated program \textsc{iSpec} \citep{2014A&A...569A.111B, 2019MNRAS.486.2075B}. Finally, we corrected for barycentric motion in the spectrum. The barycentric Julian date of the observed high-resolution spectrum is BJD 2458429.3586322.

\section{\tess\ Data Photometry}
\label{tess}

We used the 2-min cadence light curves obtained with \tess\ \citep{2015JATIS...1a4003R} from 18 July 2019 to 14 August 2019 (sector 14), 15 August 2019 to 10 September 2019 (sector 15), and 9 June 2020 to 4 July 2020 (sector 26) to search for signatures of rotational modulation and/or stellar pulsations. The light curves using pre-search data conditioning (PDC) were selected (PDCSAP\_SAP flux column in the FITS file). These are corrected for time-correlated instrumental signatures \citep{jenkins2016tess} and are good enough for our analysis.

The PDC data, as given in the FITS files downloaded from the Barbara A. Mikulski Archive for Space Telescopes (MAST)\footnote{https://mast.stsci.edu/portal/Mashup/Clients/Mast/Portal.html}, were used.
In constructing the final light curves, the mean flux was subtracted from each individual flux measurement. The result was divided by the mean flux and multiplied by 1087.5 for the conversion to millimagnitudes (mmag). Our classification of variability is based on the General Catalog of Variable Stars \citep[GCVS,][]{kazarovets2017general}.

\begin{figure}
\centering
 \includegraphics[width=\columnwidth]{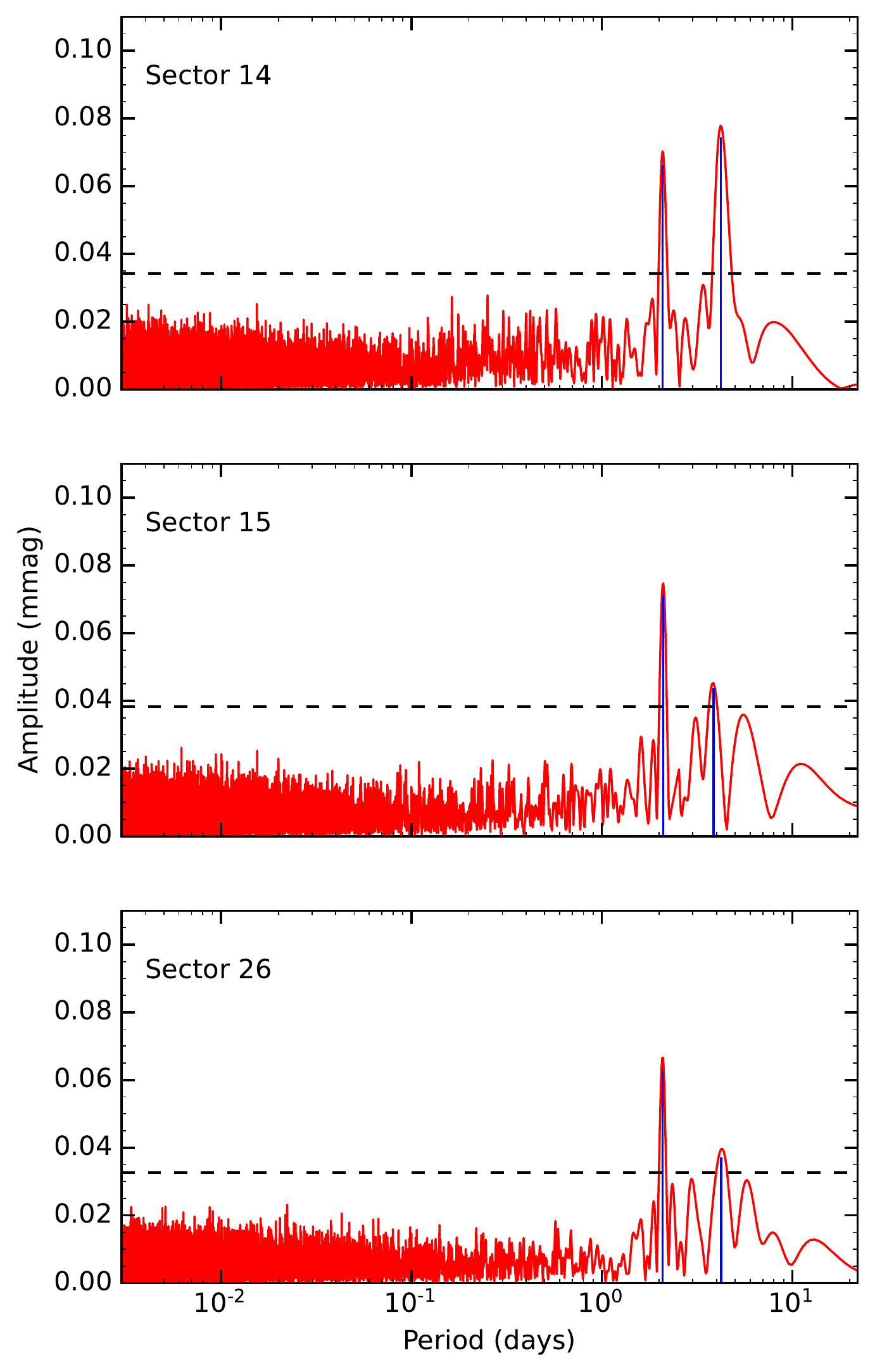}
 \caption{Linear-log periodogram distribution of spectral amplitudes, in mmag, derived from applying the Fourier transform algorithm to the light curves from each sector ({\it given in the top left corner of each panel}). The {\it blue lines} in the respective panels show the detected signals. The {\it black dashed lines} represent the threshold false alarm probability of $10^{-8}$, and the peaks above this line are considered significant signals.
 }
 \label{fig_periodogram}
\end{figure}

We calculated the amplitude SNR value for candidate signal peaks in the Fourier transform \citep{2005CoAst.146...53L} to detect significant signal peaks, using a smoothing function across the frequency spectrum to calculate the noise spectrum \citep{1993A&A...271..482B}. The amplitude of a significant peak exceeds the background noise by a factor of four or more. Simulations, as detailed in \citet{koen2010four}, support this significance criterion. Furthermore, \citet{2008MNRAS.388.1693F} noted the subjectivity of determining the noise level when calculating the SNR. We augment the SNR approach by calculating the false-alarm probability (FAP) for each peak using the independent frequency method reported in  \citet{2018ApJS..236...16V}. This criterion calculates the likelihood that a peak was caused by noise rather than an inherent signal. The lower the FAP value, the more likely it is that a given peak is real. A threshold FAP of $10^{-8}$ \citep{ 2022MNRAS.510.5854J} was adopted across all the \tess\ data sets. This threshold is also noted by \citet{2018AandA...616A..77Bowman} to be particularly significant at low frequencies. The detected signals are represented by blue lines in Fig.\,\ref{fig_periodogram} and their properties are listed in Table\,\ref{table_periodogram}. The errors listed in this table were determined using a least-square algorithm.

By visual inspection of the Fourier transforms in Fig.\,\ref{fig_periodogram}, only long-period signals (more than 1\,d) were detected. For sectors 14 and 26, the frequency values given in Table\,\ref{table_periodogram} seem to be consistent with frequencies + harmonic scenario, which suggests spot-induced rotational modulation. In the presence of the harmonic, the rotational frequency is represented by the fundamental. It is possible that the fundamental in some stars is missing or has a very low amplitude, causing the most significant period to be half that of the genuine rotational period. This could happen in stars with two almost equal-sized diametrically positioned spots (or spotted regions). Based on the Fourier transforms in Fig.\,\ref{fig_periodogram}, \target\ could be a typical example of such a case.

Low-level eclipses and rotational signals can be confused. In order to avoid this, the rotational variability identification is limited to stars with amplitudes of less than 10\,mmag. To increase the confidence of the variability type of this star, we performed wavelet, autocorrelation functions (ACFs), and composite spectrum analyses of the available \tess\ time series data. These approaches are expected to be more robust to active region evolution than the Fourier transform, which assumes an implicitly steady, sinusoidal signal.

\begin{table*}
\caption{Properties of signals detected in the \tess\ data obtained in three different sectors 14, 15, and 26. Listed is the frequency, period, amplitude, phase, probability of false alarm (log$_{10}$(FAP)), and signal-to-noise ratio (SNR) as determined from the Fourier transforms.}
\label{table_periodogram}
 \begin{tabular}{lcccccc}
\hline
\hline
\noalign{\smallskip}
\tess\ Observation & Frequency& Period & Amplitude & Phase & log$_{10}$(FAP)  & SNR \\
  Sector& (d$^{-1}$) & (d)& (mmag) & (rad) & & \\
\hline
\noalign{\smallskip}
  14 & 0.2375 $\pm$ 0.0015 & 4.211 $\pm$ 0.027 & 0.079 $\pm$ 0.006 & -1.72 $\pm$ 0.07 & -37 & 21 \\
     & 0.4798 $\pm$ 0.0016 & 2.084 $\pm$ 0.007 & 0.077 $\pm$ 0.006 & 1.41 $\pm$ 0.08 &  -32 & 19 \\
  \hline
  \noalign{\smallskip}
  15 & 0.4775 $\pm$ 0.0018 & 2.094 $\pm$ 0.008 & 0.074 $\pm$ 0.006 & -0.80 $\pm$ 0.08 & -33 & 19 \\
     & 0.2589 $\pm$ 0.0031 & 3.862 $\pm$ 0.046 & 0.043 $\pm$ 0.006 & 1.85 $\pm$ 0.13 &  -9 & 12 \\
  \hline
  \noalign{\smallskip}
  26 & 0.4803 $\pm$ 0.0017 & 2.082 $\pm$ 0.007 & 0.062 $\pm$ 0.005 & -2.69 $\pm$ 0.08 & -32 & 19 \\
     & 0.2365 $\pm$ 0.0029 & 4.228 $\pm$ 0.052 & 0.042 $\pm$ 0.005 & 2.49 $\pm$ 0.12 &  -9 & 12 \\
\hline\end{tabular}
\end{table*}

\subsection{Wavelet analysis}

Star spots could be the cause of the detected signals in the time series data. Star spots, like sunspots, are well-known tracers of stellar rotation, but their dynamic behaviour can also be used to study other phenomena like stellar magnetic activity and cycles \citep{2010Sci...329.1032G,2014A&A...562A.124M}. Sunspots change in size and location over time. Sunspots can appear or disappear at any time. If A-type star spots are comparable to sunspots, one might expect them to behave similarly. Variations in the amplitude of the frequencies are caused by variations in the size and location of starspots. To investigate the frequency change, we created time-frequency diagrams for the \tess\ data sets using the wavelet technique, which allows for a better interpretation of physical features (such as spots) before they are considered for period determination \citep{10.1175/1520-0477(1998)079-61, 2010A&A...511A..46M}.

This technique allows for the analysis of non-stationary signals for a given signal. The reference wavelet was the Morlet wavelet, which is interpreted as the convolution of a sinusoidal and a Gaussian function \citep{GOUPILLAUD198485, 1989wtfm.conf..286H}. The Morlet wavelet has several advantages: (i) it is Gaussian-shaped in the frequency domain, which minimizes ripple effects that can be misinterpreted as oscillations; (ii) the results of Morlet wavelet convolution retain the original signal's temporal resolution; and (iii) wavelet convolution is computationally efficient.

For each frequency, we calculated the correlation between the mother wavelet and the data. This was accomplished by moving the wavelet along the time axis of the light curves, producing a wavelet power spectrum (WPS). The global wavelets power spectrum (GWPS) was then generated by projecting the WPS along the period axis. Panel (c) of Figs.\,\ref{fig_wavelet1}, \ref{fig_wavelet2}, and \ref{fig_wavelet3} shows the time-frequency plots of the time series from \tess\ observation sectors 14, 15, and 26, respectively. The blue and black colours in the WPS denote low and high-power regions, respectively. By visual inspection of the GWPS's shown in panel (d) of Figs.\,\ref{fig_wavelet1}, \ref{fig_wavelet2}, and \ref{fig_wavelet3}, we report rotational periods of 4.10\,$\pm$\,0.44\,d, 3.54\,$\pm$\,0.45\,d, and 4.10\,$\pm$\,0.45\,d, respectively. Within error limits, these results are consistent with those obtained directly from Fourier transforms in Fig.\,\ref{fig_periodogram}.

\subsection{Autocorrelation functions}

The ACFs show how similar light-curves are to themselves at certain time differences \citep{2013MNRAS.432.1203M, 2014ApJS..211...24M}. Briefly, the autocorrelation function (ACF) is given by:
\begin{equation}
 ACF_\tau=\frac{1}{N}\frac{\sum^N_{i=1}(x(t_i)-\bar{x})(x(t_i-\tau)-\bar{x})}{\sigma^2},
 \label{acf1}
\end{equation} where $\tau$ is the lag time shift between the same time-series, $x(t_i)$ is the time-series value at time $t_i$, $\bar{x}$ is the temporal mean of the time series, and $\sigma^2$ is the time-series variance. The variance is given by:
\begin{equation}
 \sigma^2=\frac{\sum^N_{i=1}(x(t_i)-\bar{x})^2}{N}.
 \label{var}
\end{equation}
At time lag shift between the time series, $\tau = 0$, the ACF in Eq.\,\ref{acf1} reduces to:
\begin{equation}
 ACF_{\tau=0}=\frac{1}{N}\frac{\sum^N_{i=1}(x(t_i)-\bar{x})^2}{\sigma^2}.
 \label{acf2}
\end{equation} A comparison of Eqs.\,\ref{var} and \ref{acf2} results in $ACF_{\tau=0}=1$. When a time series contains a dominant repeated signal of the period ($P_{\rm ACF}$), probably created by the presence of spots, the pattern is expected to anti-correlate and correlate such that $ACF_{\tau=kP_{\rm ACF}/2}=-1$ and $ACF_{\tau=kP_{\rm ACF}}=1$, respectively, where $k=1, 2, 3, 4, ..., m$ and $mP_{\rm ACF}<t_N$. However, this is not the case when the whole time-series is correlated with itself due to limited overlap, for any time-lag not equal to zero. The ACF oscillates between maximum and minimum values as the patterns become correlated and anti-correlated, and the amplitude decreases as the overlap decreases.
The overall amplitude of ACFs can additionally be reduced due to variations of signals in the time series, which could be the reflection
of variations in the size of the spot (active region).
Therefore, at time lags greater than zero, the ACF resembles a displacement of an under-damped simple harmonic oscillator (uSHO) \citep{2017MNRAS.472.1618G}:
\begin{equation}
  y(\tau)=e^{-\tau/\tau_{\rm DT}}\left( {\rm A}\cos\left(\frac{2\pi \tau}{P_{\rm ACF}}\right)+{\rm B}\cos\left(\frac{4\pi \tau}{P_{\rm ACF}}\right)+y_0\right),
 \label{uSHO}
\end{equation} where
\begin{equation*}
 \tau= \Delta T \times {\rm n},
\end{equation*} $\Delta T$ is the median time difference of the light-curve, n ascends from 0 to the total number of ACFs, y($\tau$) is the ACF, $\tau_{\rm DT}$ is the decay-time scale of the ACF, $P_{\rm ACF}$ is the time lag corresponding to the first maximum of the ACF that represents the rotation period of the star,
and A, B, and $ y_0$ that do not represent any physical stellar properties but are constants that are needed in the fit of an uSHO.

Using Eq.\,\ref{acf1}, we calculated ACFs of the time-series at different time-lags as represented in panel\,(e) of Figs.\,\ref{fig_wavelet1}, \ref{fig_wavelet2}, and \ref{fig_wavelet3}. The first dominant peak in the ACF was selected as the $P_{\rm ACF}$. The obtained values of $P_{\rm ACF}$ are 4.1\,d, 2\,d, and 1.98\,d, for sectors 14, 15, and 26, respectively. In panel (e) of Fig.\,\ref{fig_wavelet1}, the ACF shows sub-peaks at  $\tau=kP_{\rm ACF}/2$. The sub-peaks in panel (e) of Figs.\,\ref{fig_wavelet2} and \ref{fig_wavelet3} are of comparable strength to the major peaks. The strength of the sub-peaks increases due to the dominance of secondary signals (perhaps harmonics) as shown in panels (b), (c), and (d) of Figs.\,\ref{fig_wavelet2} and \ref{fig_wavelet3}. This is thought to occur when slightly weaker spotted regions are diametrically opposite to the dominant spotted region as earlier mentioned in Section\,\ref{tess}.  This means that the obtained $P_{\rm ACF}$ values for sectors 15 and 26 are half the genuine values.

\subsection{Composite spectrum}
The ACF and the GWPS are sensitive to different issues in the light curve. When we combine the two, we can find periods that are intrinsic to the star. The composite spectrum (CS) combines the two preceding methods (wavelet and ACF) \citep{10.1093/mnras/stv2622,2017A&A...605A.111C}.

We used an exponentially decreasing function to fit the smoothed ACF. To obtain the normalised ACF, the fit was subtracted from the smoothed ACF.  We calculated the CS by multiplying the normalized ACF and GWPS. This increases the height of the peaks in both curves while decreasing the height of the peaks in one of the two.  The CS is very sensitive to the periods detected in both ACF and GWPS, which makes it reliable up to about 95\% of the time \citep{2015MNRAS.450.3211A}.

\begin{figure*}
\centering
 \includegraphics[width=\textwidth]{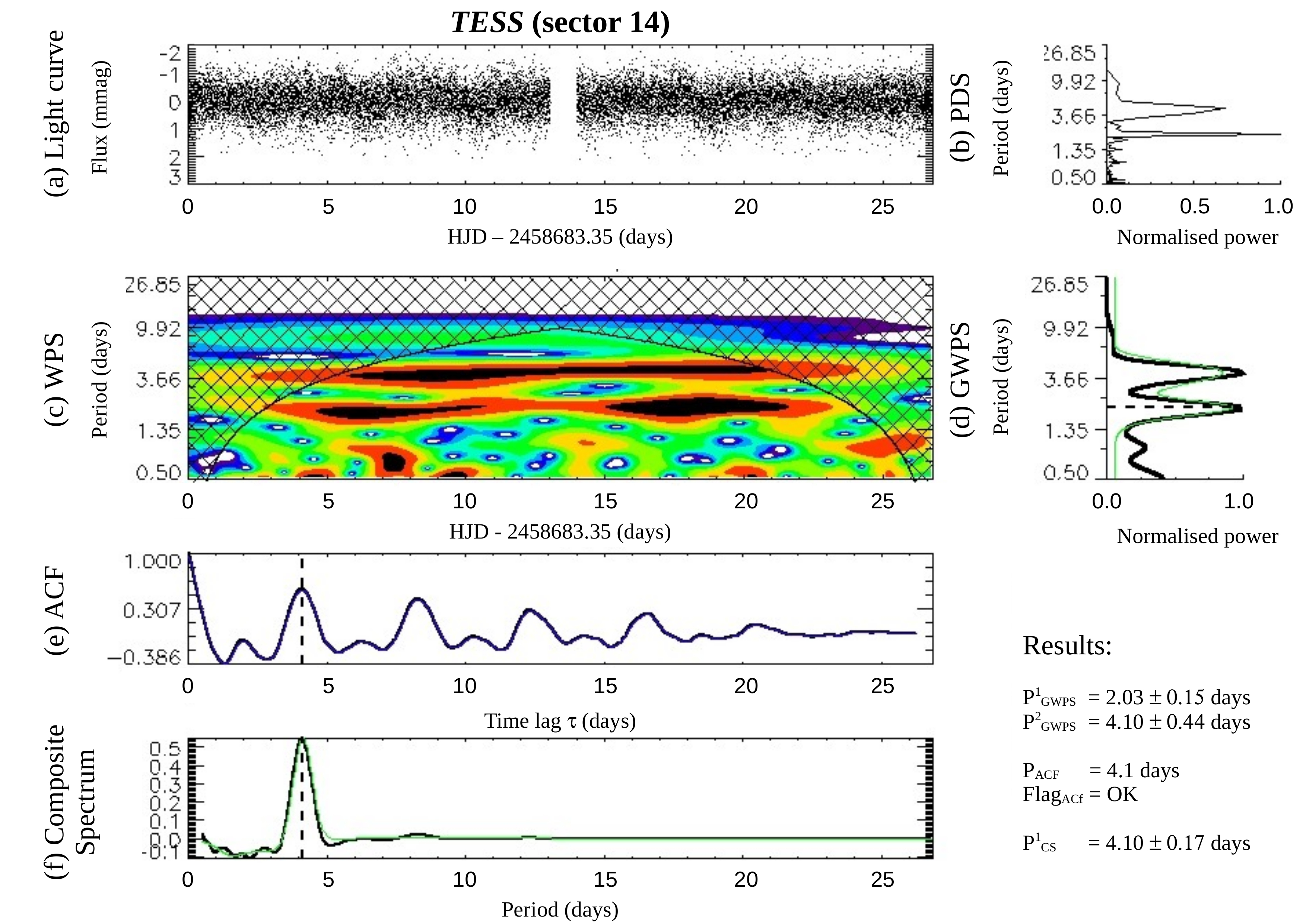}
 \caption{Search for the rotation period of \target\ based on the \tess\ data from sector 14. Panel (a) shows the light curve while the power density spectrum (PDS) as a function of the period between 0.5 and 27\,d is given in panel (b).  The wavelet power spectrum (WPS) computed using a Morlet wavelet between 0.5 and 27\,d on a logarithmic scale is shown in  panel (c)  and the associated global wavelet power spectrum (GWPS) is given in panel (d).  The colours black and blue represent high and low power, respectively. The autocorrelation function (ACF) of the full light curve and the composite spectrum (CS) \citep{10.1093/mnras/stv2622, 2017A&A...605A.111C}, plotted between 0 and 27\,d, are presented in panels (e) and (f), respectively. The cone of influence corresponding to the unreliable results is represented by the black-crossed area in the WPS. The automatically detected rotational period estimates are denoted by the black dashed lines. A summary of the results is given in the bottom right corner. The quality flag Flag$_{\rm ACF}$ indicates whether the selected $P_{\rm ACF}$ corresponds to the dominant of the regularly spaced peaks of the ACF.}
 \label{fig_wavelet1}
\end{figure*}

\begin{figure*}
\centering
 \includegraphics[width=\textwidth]{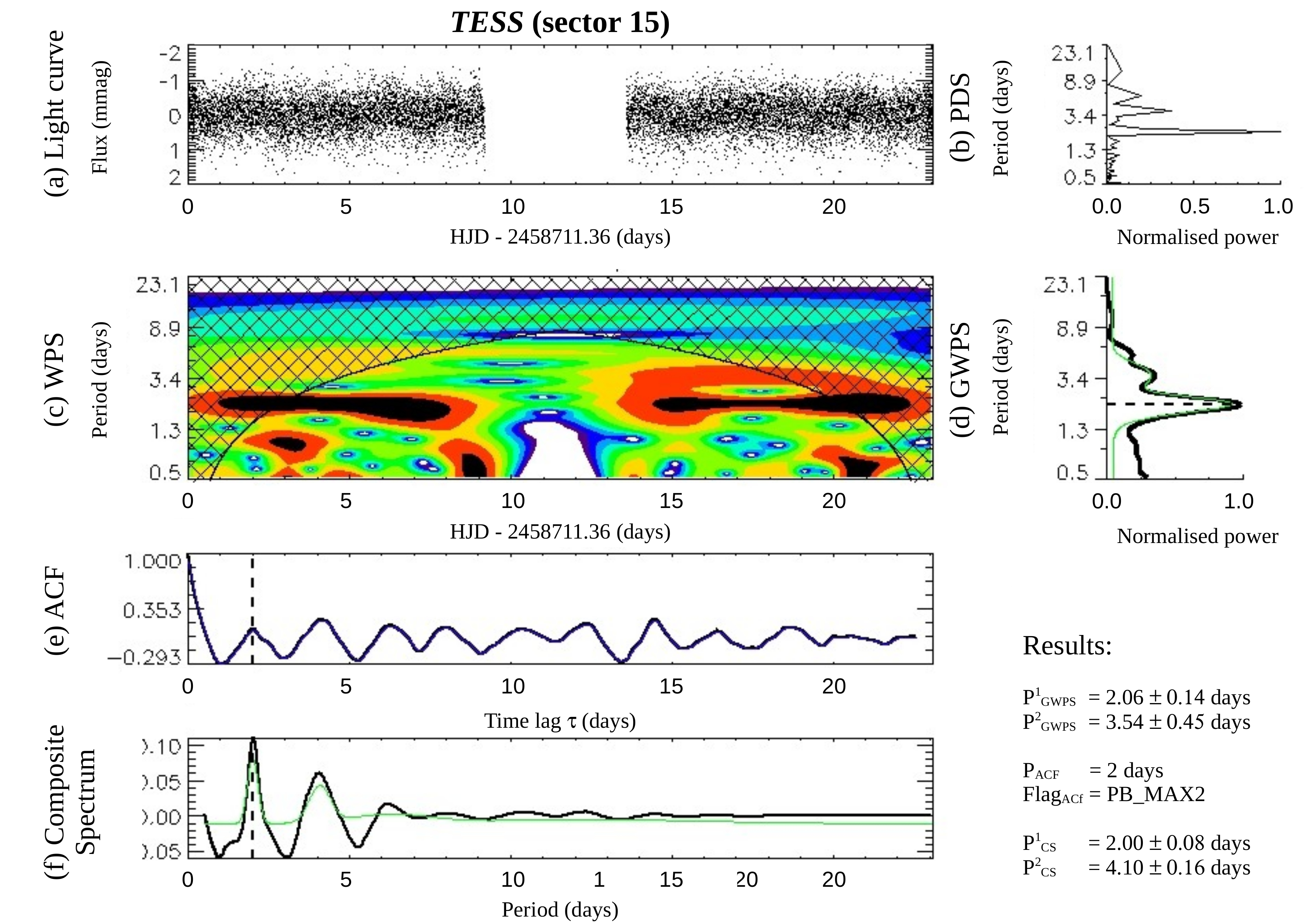}
 \caption{Same as Fig.\,\ref{fig_wavelet1} but for the \tess\ data from sector 15. }
 \label{fig_wavelet2}
\end{figure*}

\begin{figure*}
\centering
\includegraphics[width=\textwidth]{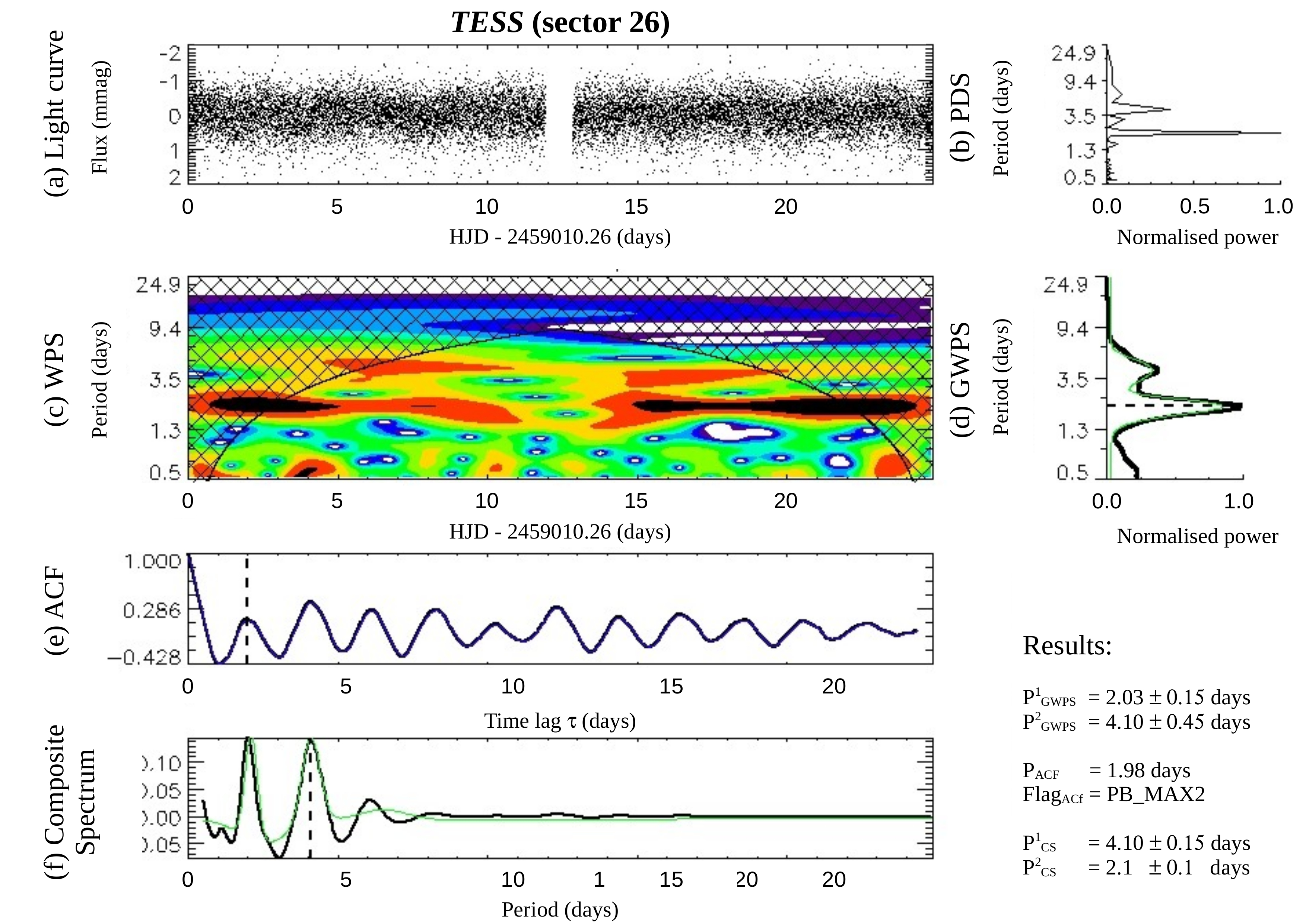}
 \caption{Same as Fig.\,\ref{fig_wavelet1} but for the \tess\ data from sector 26. }
 \label{fig_wavelet3}
\end{figure*}

We calculated the period, $P_{\rm CS}$, by fitting the peaks in the CS with Gaussian functions. The central period of the function corresponding to the highest peak was taken as the $P_{\rm CS}$. The associated uncertainty corresponds to the peak's half width at half maximum (HWHM). By visual inspection of panel (f) of Figs.\,\ref{fig_wavelet1}, \ref{fig_wavelet2}, and \ref{fig_wavelet3}, we report rotational periods of 4.10\,$\pm$\,0.17\,d, 4.10\,$\pm$\,0.16\,d, and 4.10\,$\pm$\,0.15\,d, respectively. The only difference among the $P_{\rm CS}$ values is the uncertainty.

The results of {\it TESS} photometry investigations show no evidence of pulsational variability, as previously proposed by \citet{2019MNRAS.484.2530C}. Following the criterion by \citet{2013MNRAS.432.1203M}, \citet{10.1093/mnras/stv2622}, and \citet{2017A&A...605A.111C}, the observed signals are rotational. Based on the strengths of the methods \citep{2017A&A...605A.111C}, we take $P_{\rm CS}$ as the ultimate rotational period for HD\,180347. The amplitude of the rotational signal indicates the spot size, which gives a clue about the magnetic field's strength. The rotational signal amplitudes range from 0.07 to 0.08\,mmag, which is much larger than the average amplitude (0.02\,mmag) of A and Am stars with "hump and spike" features in their Fourier transforms investigated by \citet{2020MNRAS.492.3143T}. This means that HD\,180347 has spots that are about four times larger and so have greater magnetic fields. Magnetic fields are vital in stabilizing the material, which allows for atomic diffusion. A spectropolarimetric analysis of this star is required to determine the strength of its magnetic field. This, however, is beyond
the scope of this research.

\section{Spectral classification}
\label{class}

We used the MK classification system \citep{1943assw.book.....M,2009ssc..book.....G} to perform a spectral classification analysis on \target.
It can be used to determine the chemical peculiarity of a star \citep{2009ssc..book.....G}. The spectral type and luminosity class are determined by comparing the observed spectra to well-known standards while accounting for crucial hydrogen and metal lines. We used the high-resolution HERMES spectrum to do the spectral classification of \target.


The HERMES spectrum of \target\ was classified with the help of the code \textsc{Mkclass}\footnote{http://www.appstate.edu/~grayro/mkclass/} \citep{2014AJ....147...80G}
in combination with the standard libraries given by \citet{2003AJ....126.2048G}. The spectra of this library were obtained with the 0.8-m telescope of the Dark Sky Observatory (DSO) in the northwest of North Carolina (USA) by using the Gray/Miller classification spectrograph with a grating having either 600 or 1200 grooves/mm$^{-1}$. The standard spectra span the violet-green wavelength region at a resolution range of 0.18\,--0.36\,nm / 2 pixels \citep{2014AJ....147...80G}.

Our HERMES spectrum is not observed with the same spectrograph/grating combination as the spectra of the standard library. We, therefore, truncated its wavelength region, re-binned it, and convoluted the spectrum with a Gaussian of appropriate full width at half maximum of 0.16\,nm to match the specifications
of the standards as closely as possible.

\textsc{Mkclass} uses the metric-distance technique \citep{1994ASPC...60..312L}, which is based on a weighted least-square comparison of the program spectrum with that of the MK standard stars
\citep{2014AJ....147...80G}, to determine the spectral type based on (i) hydrogen lines ($\rm H\gamma$ and $\rm H\sigma$), (ii) metal lines, and (iii) the Ca\,\textsc{ii}\,K-line. For a chemically normal star, this should lead to the same results while different spectral types in these three regions are expected in the case of a chemically peculiar star \citep{2009ssc..book.....G}. With this method, we found a spectral type of kA1hA8mA8, confirming the classification of \target\ as an Am star, first reported by \citet{1985AJ.....90..341B}.

\section{Fundamental stellar parameters}
\label{sect:fund_par}

High-resolution spectroscopy is a robust method to derive accurate values of the basic stellar parameters like the effective temperature (\teff), surface gravity (\logg), and metallicity (\mh) (Section\,\ref{sect:deriv_par}). Once they are known, they can be used to calculate values of additional fundamental stellar parameters (Section\,\ref{sect:calc_par}).

\subsection{Derived stellar parameters}
\label{sect:deriv_par}

To achieve convergence during synthesis, knowledge of a good initial guess for the basic stellar parameters is critical. Therefore, we derived \teff, \logg, and \mh\ in two steps.

\begin{figure*}
\centering
 \includegraphics[width=0.7\textwidth]{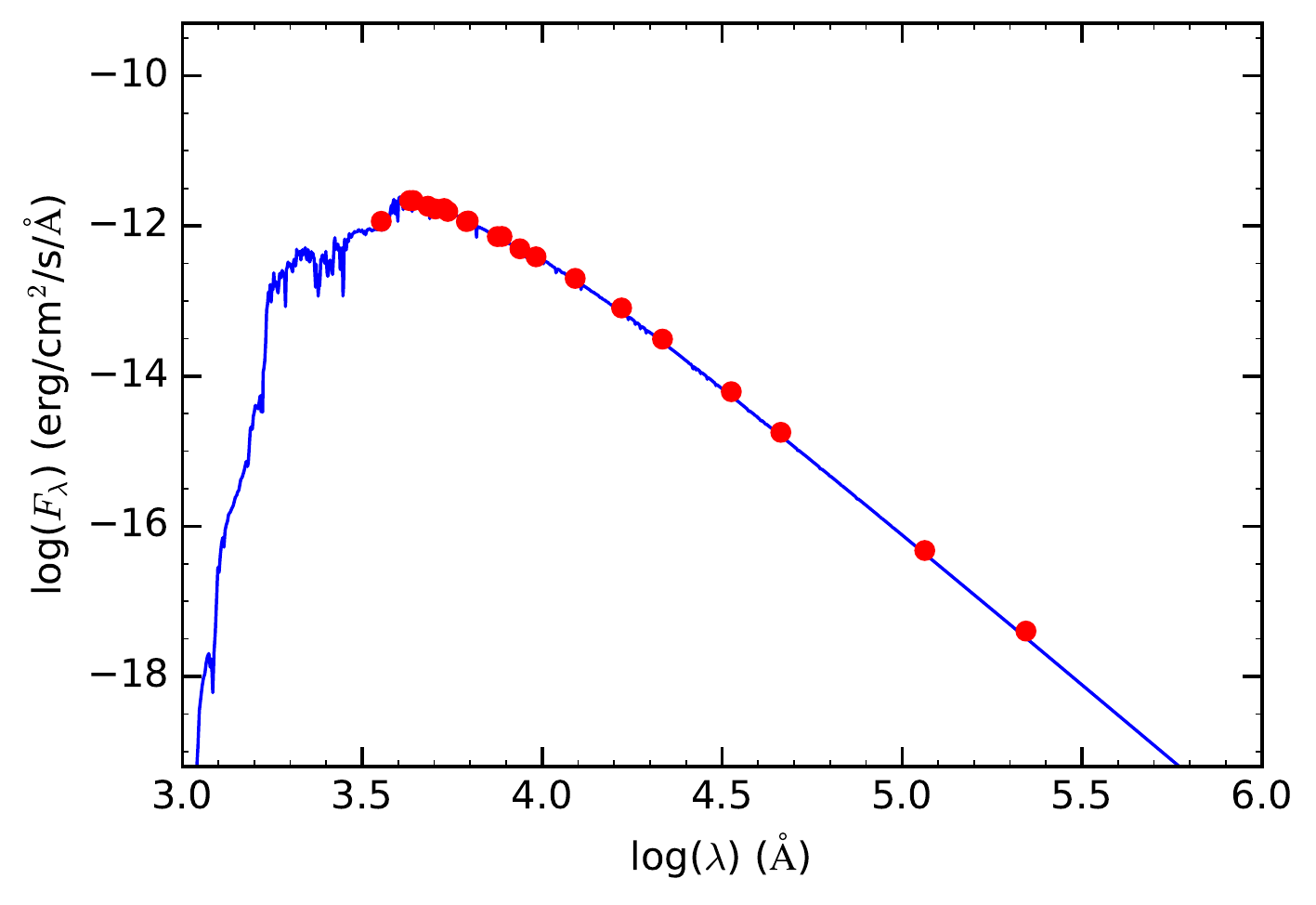}
 \caption{The red-filled circles represent the SED of \target. The blue solid line represents the best fit to the data obtained using the \textsc{vosa} tool.}
 \label{sed_fig}
\end{figure*}

In the first step, the spectral energy distribution (SED) of \target\ was used to obtain preliminary values for \teff, \logg, and \mh\ \citep[cf.][]{2021MNRAS.504.5528T}. Using the \textsc{vosa}\footnote{http://svo2.cab.inta-csic.es/theory/vosa/} \citep{2008A&A...492..277B} tool and the ATLAS9 Kurucz ODFNEW/NOVER models \citep{2003IAUS..210P.A20C}, we performed a least-square fit to the SED to obtain \teff, \logg, and [M/H] as 7750\,$\pm$\,250\,K, 4.0\,$\pm$\,0.5\,cm\,s$^{-2}$, and 0.5\,$\pm$\,0.30\,dex, respectively. The uncertainties in the parameters are the propagated uncertainties in the fit parameter calculation. We emphasise that these SED results only served the purpose of being first guesses for spectroscopic analysis. The best model is represented by a blue line in Fig.\,\ref{sed_fig}.

In the second step, the HERMES spectrum was used to derive the radial velocity (\vrad; Section\,\ref{sect:vrad}), \vsini\ (Section\,\ref{sect:vsini}), and the microturbulent velocity ($\xi$) and final values of \teff, \logg, and \mh\ (Section\,\ref{sect:basic}).

\subsubsection{Radial velocity}
\label{sect:vrad}

We calculated the radial velocity \vrad\ by computing the cross-correlation function (CCF) with various pre-selected masks created from line lists using the code \textsc{iSpec}\footnote{https://www.blancocuaresma.com/s/iSpec}. We found a \vrad\ value of 9.3\,$\pm$\,0.3\,\kms\ corresponding to the time that the HERMES spectrum was observed (BJD 2458429.3586322).

\subsubsection{Projected rotational velocity}
\label{sect:vsini}

Using the initial values from the SED analysis,
\vsini\ was calculated by comparing the observed spectrum to a grid of synthetic spectra. The least-square method was used in this comparison, which was based on the {\sc minuit} minimization software, which is included in the {\sc girfit} package \citep{2006A&A...451.1053F}. We interpolated the spectrum in a grid of stellar fluxes computed using plane-parallel ATLAS9\footnote{http://www.stsci.edu/hst/observatory/crds/castelli\_kurucz\_atlas.html} model atmospheres \citep{2003IAUS..210P.A20C} for \teff, \logg, and \vrad\ values of 7750\,K, 4.0\,cm\,s$^{-2}$, and 9.3\,\kms, respectively, and varied \vsini\ values in the range 0\,--\,100\,\kms\ with steps of 1\,\kms.

Based on the initial prediction of \teff\ (7750\,K) from SED, we chose the Mg\,\textsc{i} triplet region (516 - 519\,nm) for \vsini\ determination. In addition to other metal lines, stars with \teff\ less than 9000\,K, Mg\,\textsc{i} triplet is sensitive to the \vsini\ \citep[cf.][]{catanzaro2015, 2019MNRAS.484.2530C, 2021MNRAS.504.5528T}. In this spectral region, Mg\,\textsc{i} triplet lines predominate; nevertheless, additional metal lines such as iron (516.227, 516.541, 516.628, 517.16\,nm), nickel (517.656\,nm), and titanium (518.59, 518.869\,nm) are also present. With this method, the resulting value for \vsini\ is 14\,$\pm$\,2\,\kms.

\subsubsection{Effective temperature, surface gravity, metallicity, and microturbulent velocity}
\label{sect:basic}

We determined \teff, \logg, \mh, and $\xi$ by synthesizing stellar spectra with the MOOG radiative transfer code\footnote{https://www.as.utexas.edu/~chris/moog.html} \citep{2012ascl.soft02009S}. We used the ATLAS9 model atmospheres, the Vienna Atomic Line Database (VALD) line list \citep{1999A&AS..138..119K}, and the solar abundances of \citet{2009ARA&A..47..481A}, all of which were combined in the integrated software package \textsc{iSpec}. The Balmer lines are \teff\ sensitive but lose \logg\ sensitivity for stars with \teff\ less than 8000\,K. \teff\ and \logg\ were estimated from the hydrogen line profiles and Fe\,\textsc{i}/Fe\,\textsc{ii} lines, respectively. \mh\ was determined from all available lines, with Fe, Ca, and Ti lines dominating. Since \vsini\ is low, the $\xi$ was determined by fitting Fe and Ti\,\textsc{ii} lines and its initial guess was calculated from the relation \citep{gebran14},
\begin{equation}
 \xi=3.31 \times \exp\left[-\left(\log\left(\frac{T_{\rm eff}}{8071.03}\right)^2/0.01045\right)\right].
\end{equation}
The resulting values are 7740\,$\pm$\,170\,K for \teff, 3.98\,$\pm$\,0.12\,cm\,s$^{-2}$ for \logg, 0.11\,$\pm$\,0.08\,dex for \mh, and 3.81\,$\pm$\,0.12\,\kms\ for $\xi$. Uncertainties in parameters were calculated as the change in parameter values that raises $\chi^2$ by one \citep{1976ApJ...208..177L}.

\begin{figure*}
\centering
 \includegraphics[width=0.7\textwidth]{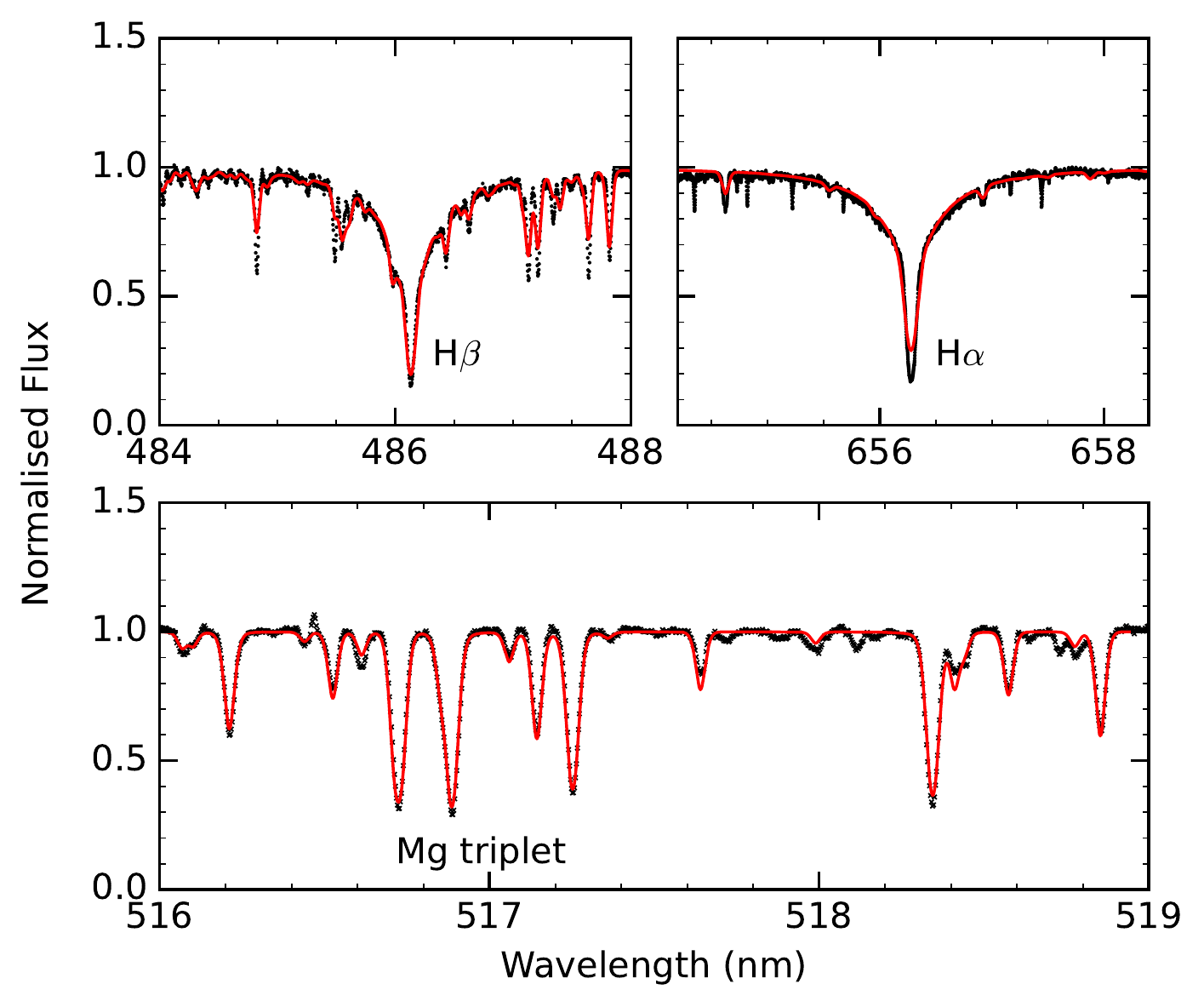}
 \caption{The \hbeta\ ({\it top left}), Mg\,\textsc{i} triplet ({\it bottom}), and \halpha\ ({\it top right}) line regions for the target star. The observed ({\it black}) and synthetic spectra ({\it red}) are shown. The synthetic spectrum was obtained with LTE consideration. The majority of the narrow features in the observed spectrum that the models do not fit are telluric lines.}
 \label{hb_fig}
\end{figure*}

\smallskip
Fig.\,\ref{hb_fig} shows the \hbeta\ ({\it top left}), Mg\,\textsc{i} triplet ({\it bottom}), and \halpha\ ({\it top right}) line regions for the HERMES spectrum of \target\ ({\it black}) and the synthetic ({\it red}) spectra computed with the final values of the atmospheric parameters as listed in the top part of Table\,\ref{tab:sect:fund_par}. Our results, within the error bounds, accord with some of those from prior studies, as indicated in Table\,\ref{tab:sect:fund_par}. The literature values that differ from our results are shown in italics in this table.

\begin{table}
\centering
\caption{Overview of the values of fundamental parameters of \target\ resulting from this study  and those available in the literature. The top part lists the parameters directly obtained from the HERMES spectrum while the bottom part gives those derived from them.  Within the error limits, the literature values that disagree with our results are provided in italics.}
\label{tab:sect:fund_par}
\begin{tabular}{llll}
\hline
\hline
\noalign{\smallskip}
Parameter& Value & Error & Reference\\
\noalign{\smallskip}
\hline
\noalign{\smallskip}
\teff\ (K) & 7740  & 170  & This study\\
           &7685   & & \citet{2012MNRAS.427..343M}\\
           &7900   & 140  & \citet{catanzaro2015}\\
           & 7709  & 264  & \citet{2017ApJS..229...30M} \\
           & {\it 7522}  & 97  & \citet{2018AA...616A...8A} \\
           &7699   & 269  & \citet{10.1093/mnras/stz590}\\
           &7600   & 125 & \citet{2019MNRAS.484.2530C}\\
           & {\it 7544} & 97 & \citet{2022arXiv220606138A}\\
\noalign{\smallskip}
\logg\ (\,cm\,s$^{-2}$)         &   3.98 &   0.12 & This study\\
           & {\it 3.85}  & 0.07 & \citet{catanzaro2015} \\
           & 3.99  & 0.19 & \citet{2017ApJS..229...30M} \\
           & {\it 4.11}  & 0.09 & \citet{10.1093/mnras/stz590} \\
           & 4.0   & 0.25 & \citet{2019MNRAS.484.2530C}\\
           & {\it 4.13} & & \citet{2022arXiv220606138A}\\
\noalign{\smallskip}
\mh\ (dex)           &   0.11 &   0.08 &This study\\
\noalign{\smallskip}
$\xi$ (\kms)         &   3.81 &   0.12 &This study\\
           & {\it 4.7}     & 0.4 & \citet{catanzaro2015} \\
\noalign{\smallskip}
\vsini\ (\kms)       &     14 &      2 &This study\\
           & 12      & 1 & \citet{catanzaro2015}\\
           & {\it 11}      & 1 & \citet{2019MNRAS.484.2530C}\\
           & 14.64  &   & \citet{2020AJ....160..120J}\\
\noalign{\smallskip}
\vrad\ (\kms)        &    9.3 &    0.3 &This study\\
           & {\it 5.1}      & 0.1 & \citet{2019MNRAS.484.2530C}\\
           & 9.30  &  0.01 & \citet{2020AJ....160..120J}\\
           & {\it 5.2}      & 1 & \citet{2022arXiv220605486B} \\
\hline
\noalign{\smallskip}
\luminosity          &  0.993 &  0.030 &This study\\
            & 0.97  &   & \citet{2018AA...616A...8A} \\
            & {\it 0.88}      &  & \citet{2019MNRAS.484.2530C}\\
\noalign{\smallskip}
\radius\ (R$_\odot$) &   1.75 &   0.14 &This study\\
         & {\it 2.19}  & 0.56  & \citet{2017ApJS..229...30M} \\
         & 1.80  & 0.05  & \citet{2018AA...616A...8A} \\
         &  1.83  & 0.14 & \citet{10.1093/mnras/stz590} \\
\noalign{\smallskip}
$v_{\rm eq}$ (\kms)  &     22 &      1 &This study\\
\noalign{\smallskip}
$i$ (\textdegree)    &     40 &      5 &This study\\
\noalign{\smallskip}
\mass\ (M$_\odot$)   &   1.740 &   0.023 &This study\\
        & 1.71  & 0.26  & \citet{2017ApJS..229...30M} \\
        & 1.58  & 0.25 & \citet{10.1093/mnras/stz590} \\
\noalign{\smallskip}
age (Gyr)            &   0.85 &   0.18 &This study\\
            &  0.14 -- 1.1  & & \citet{2019MNRAS.484.2530C}\\
\hline
\end{tabular}\\
\end{table}
\subsection{Calculated stellar parameters}
\label{sect:calc_par}

\subsubsection{Luminosity and radius}
\label{sect:lum_radius}

Based on the standard technique, we determined the luminosity (\luminosity) \citep[cf.][]{2021MNRAS.504.5528T}. We determined the reddening parameter $E{\rm (B-V)}$ from 3D models \citep{bayestar, 2019ApJ...887...93G} using the GAIA parallaxes \citep{2018A&A...616A...4E} and the stellar galactic coordinates from the SIMBAD database{\footnote{https://simbad.u-strasbg.fr/simbad/}} \citep{2000A&AS..143....9W} and found a value of 0.0173\,$\pm$\,0.0015\,mag. Using the temperature-dependent function by \citet{1996ApJ...469..355F} revised by \citet{Torres_2010}, the bolometric correction (BC) was calculated. We computed the absolute magnitude ($M_{\rm v}$) using GAIA parallaxes \citep{2018A&A...616A...4E}. Uncertainties in V-band indices, parallax, and \teff\ contribute to the uncertainties in \luminosity. We used the method described in \citet{2021MNRAS.504.5528T} to calculate the stellar radius (\radius) from the Stefan--Boltzmann law \citep{1884AnP...258..291B, 2015PhST..165a4027P, Montambaux2018}. The procedure produced a $M_{\rm v}$ value of 2.252\,$\pm$\,0.083\,mag, a BC value of 0.0283\,$\pm$\,0.0055\,mag, a \luminosity\ value of 0.993\,$\pm$\,0.030, and a \radius\ value of 1.75\,$\pm$\,0.14\,R$_\odot$. Within the error limits, our results are in agreement with those reported in \citet{2018AA...616A...8A} as shown in Table\,\ref{tab:sect:fund_par}.

\subsubsection{Equatorial rotational velocity and inclination angle}
\label{sect:incl}

Using the rotational period (4.10\,$\pm$\,0.17\,days) obtained from the \tess\ data and the value of \radius, the equatorial rotational velocity $v_{\rm eq}$ was calculated \citep[cf.][]{2020MNRAS.492.3143T}. We constrained the inclination angle ($i$) from the relationship between \vsini\ and the equatorial rotational velocity ($v_{\rm eq}$). We report the values of $v_{\rm eq}$ and $i$ to be 22\,$\pm$\,1\,\kms\ and 40\,$\pm$\,5\textdegree, respectively.

\subsubsection{Stellar mass and age}
\label{sect:mass_age}

\begin{figure*}
\centering
 \includegraphics[width=0.7\textwidth]{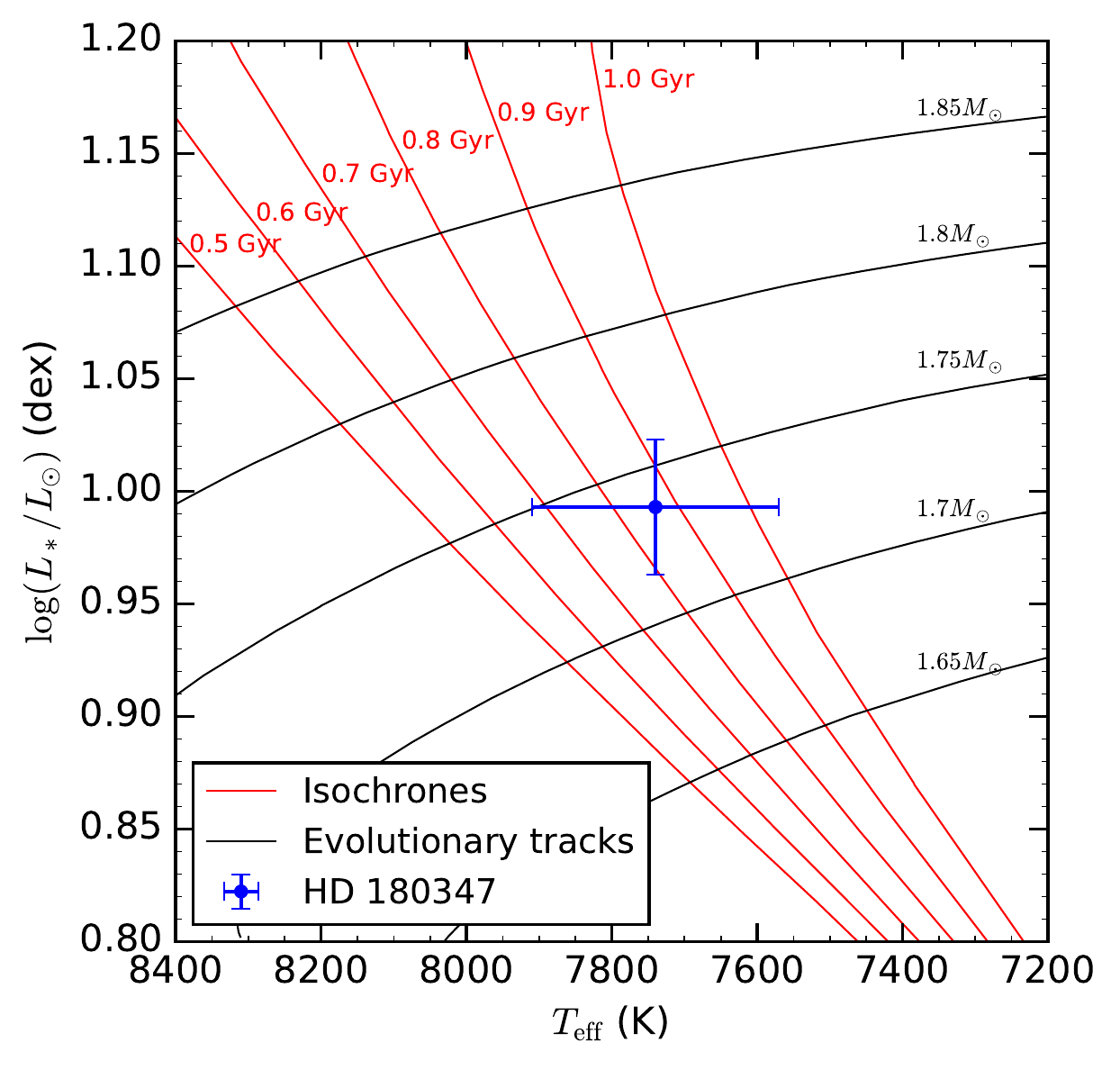}
 \caption{The HR diagram showing the position of \target\ relative to a number of the PARSEC\,1.2 evolutionary tracks ({\it black solid lines}) and isochrones ({\it red solid lines}) \citep{2012MNRAS.427..127B}.}
 \label{hrd_fig}
\end{figure*}

Using the \teff\ and \luminosity\ values derived in the previous sections, the stellar mass (\mass) and age were determined by interpolating the PARSEC\,1.2 evolutionary tracks and isochrones \citep{2012MNRAS.427..127B}, respectively. The grids of evolutionary tracks span a mass range of 1.5\,--\,2.5\,M$_\odot$ while the isochrones have ages ranging between 0.2 and 1\,Gyr. The results for \mass\ and age are 1.740\,$\pm$\,0.023\,M$_\odot$ and 0.85\,$\pm$\,0.18\,Gyr, respectively. Fig.\,\ref{hrd_fig} represents a Hertzsprung--Russell (HR) diagram showing the position of \target\ relative to a number of the PARSEC\,1.2 evolutionary tracks and isochrones. The values of the fundamental parameters derived in this section are listed in the bottom part of Table\,\ref{tab:sect:fund_par}.

\section{Individual chemical abundances}
\label{sect:abun}

With a high-resolution HERMES spectrum with an SNR\,$\sim$100 at hand, it is possible to derive the abundances of chemical elements with a sufficient number of detectable absorption lines in the observed spectrum. We started with an abundance analysis for all the elements assuming local thermodynamic equilibrium (LTE; Section\,\ref{sect:abun_lte}). Afterwards, for 13 elements, we investigated the effects of the non-local thermodynamic equilibrium (NLTE) line formation on their derived abundances (Section\,\ref{sect:abun_nlte}).

\subsection{LTE abundance analysis}
\label{sect:abun_lte}

The individual chemical abundances were calculated via direct fitting of theoretical profiles of individual spectral lines using the \vsini\ and \vrad\ values derived in Section\,\ref{sect:deriv_par}. We used the \textsc{SynthV\_NLTE} code \citep{2019ASPC..518..247T} in combination with a grid of pre-computed atmospheric models from the LLmodels package \citep{2004A&A...428..993S}. The calculations were carried out using the IDL visualization program \textsc{BinMag6}\footnote{https://www.astro.uu.se/~oleg/binmag.html} \citep{2018ascl.soft05015K}. LTE was assumed during spectrum synthesis. The line lists and atomic parameters were extracted from the 3D release of the Vienna Atomic Line Database (VALD3; \citealt{Ryabchikova_2015}).

\begin{figure*}
\centering
 \includegraphics[width=\textwidth]{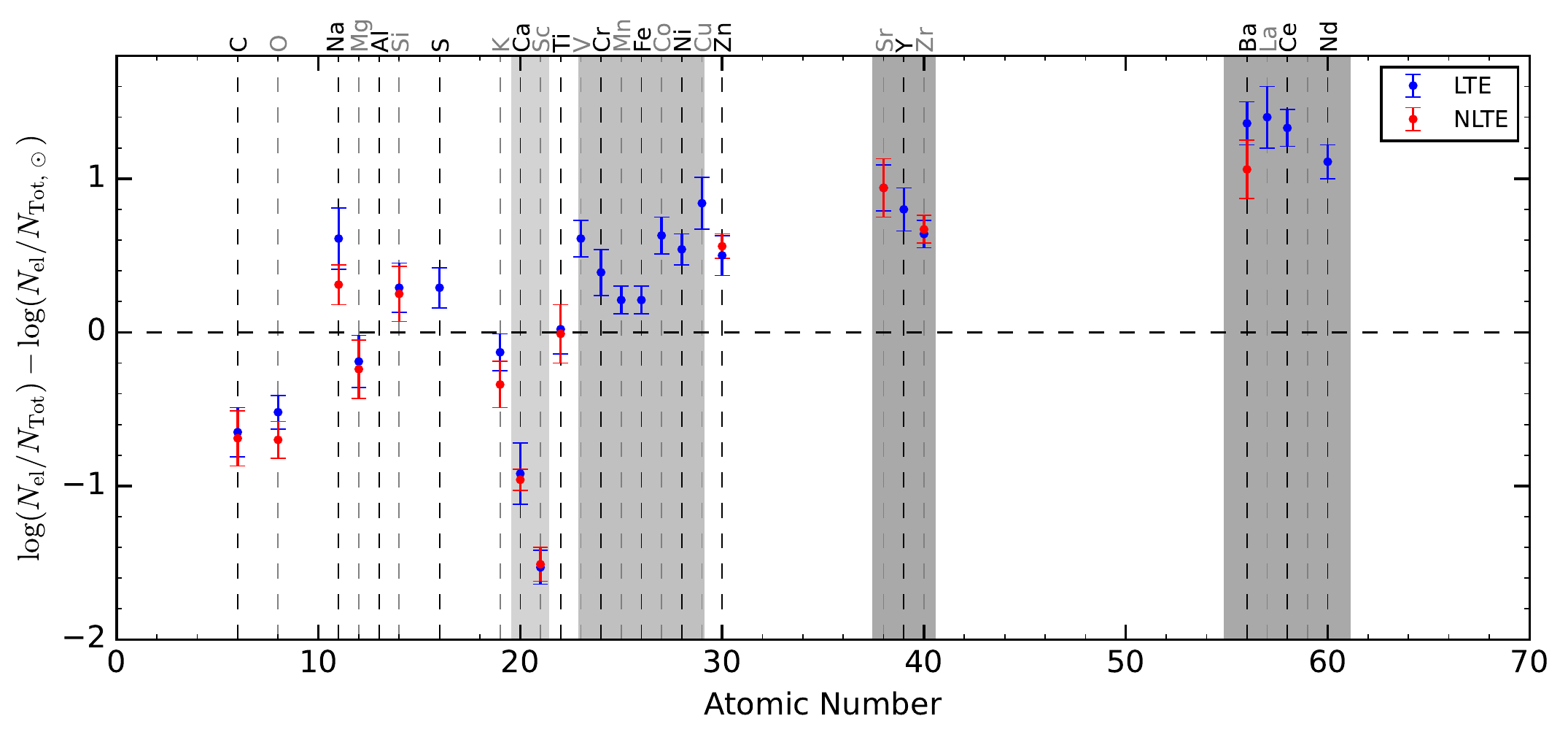}
 \caption{The individual chemical abundance pattern for \target.
 The horizontal dashed line indicates solar abundances
 \citep{2009ARA&A..47..481A}. The light elements (Ca and Sc) and heavy elements (Fe, Co, Ni, Cu, Sr, Y, Zr, Ba, La, Ce, Pr, and Nd),
 important for the classification of Am stars, are highlighted with a gray background. The {\it blue} and {\it red} symbols represent the LTE and NLTE abundances, respectively. }
 \label{fig:abundances}
\end{figure*}

We used a $\chi^2$ minimization of the difference between the observed and synthetic spectrum to derive the abundances for each individual line present in the wavelength interval. Table\,\ref{table:chem} shows the average individual chemical abundances and their uncertainties expressed as log$(N_{\rm el}/N_{\rm Tot})$.  The abundance pattern, in relation to solar abundances \citep{2009ARA&A..47..481A}, is shown in Fig.\,\ref{fig:abundances}. The uncertainties in abundances result from a combined dependence of
the errors on \teff, \logg, \vsini, $\xi$, the position of the continuum of the normalised HERMES spectrum, and the accuracy of the oscillator strengths ($\log({\rm gf})$) of the lines considered in our analysis. We should be cautious about the abundances of elements whose lines are present in less than three spectral lines because they were calculated from a small number of lines.

\begin{table}
\centering
\caption{Results of the abundance analysis of \target\ based on the observed HERMES spectrum and with LTE consideration. Listed in column 2 are the individual chemical abundances inferred for our target star and column 3 gives the solar abundances \citep{2009ARA&A..47..481A}. The number of spectral lines from which the abundances were derived is given between brackets.}
\label{table:chem}
\setlength{\tabcolsep}{2pt}
\begin{tabular}{p{.6in}p{.3in}p{.05in}p{.2in}p{.45in}c}
\hline
\hline
\noalign{\smallskip}
Element &\multicolumn{4}{c}{$\log(N_{\rm el}/N_{\rm Tot})_*$~~~~~~~~}&$\log(N_{\rm el}/N_{\rm Tot})_\odot$\\
\noalign{\smallskip}
\hline
  C  & -4.26 & $\pm$ & 0.16 & (7) &  -3.61\\
  O  & -3.87 & $\pm$ & 0.11 & (3) &  -3.35\\
  Na & -5.19 & $\pm$ & 0.20 & (6) &  -5.80\\
  Mg & -4.63 & $\pm$ & 0.17 &(14) &  -4.44\\
  Si & -4.24 & $\pm$ & 0.16 & (43) &  -4.53\\
  S  & -4.63 & $\pm$ & 0.13 & (22) &  -4.92\\
  K  & -7.14 & $\pm$ & 0.12 & (1) &  -7.01\\
  Ca & -6.62 & $\pm$ & 0.20 &(28) &  -5.70\\
  Sc & -10.42 & $\pm$ & 0.11 & (4) &  -8.89\\
  Ti & -7.07 & $\pm$ & 0.16 &(102) &  -7.09\\
  V  & -7.50 & $\pm$ & 0.12 & (36) &  -8.11\\
  Cr & -6.01 & $\pm$ & 0.15 &(127) &  -6.40\\
  Mn & -6.40 & $\pm$ & 0.09 & (57) &  -6.61\\
  Fe & -4.33 & $\pm$ & 0.09 &(368) &  -4.54\\
  Co & -6.42 & $\pm$ & 0.12 &(20) &  -7.05\\
  Ni & -5.28 & $\pm$ & 0.10 &(121) &  -5.82\\
  Cu & -7.01 & $\pm$ & 0.17 &(7) &  -7.85\\
  Zn & -6.98 & $\pm$ & 0.13 & (2) &  -7.48\\
  Sr & -8.23 & $\pm$ & 0.15 & (4) &  -9.17\\
  Y  & -9.03 & $\pm$ & 0.14 & (29) &  -9.83\\
  Zr & -8.87 & $\pm$ & 0.09 &(4) &  -9.51\\
  Ba & -8.49 & $\pm$ & 0.10 & (4) &  -9.85\\
  La & -9.54 & $\pm$ & 0.20 & (28) & -10.94\\
  Ce & -9.13 & $\pm$ & 0.12 & (32) & -10.46\\
  Nd & -9.51 & $\pm$ & 0.11 & (18) & -10.62\\
\hline
\end{tabular}
\end{table}

\begin{figure*}
\centering
 \includegraphics[width=0.7\textwidth]{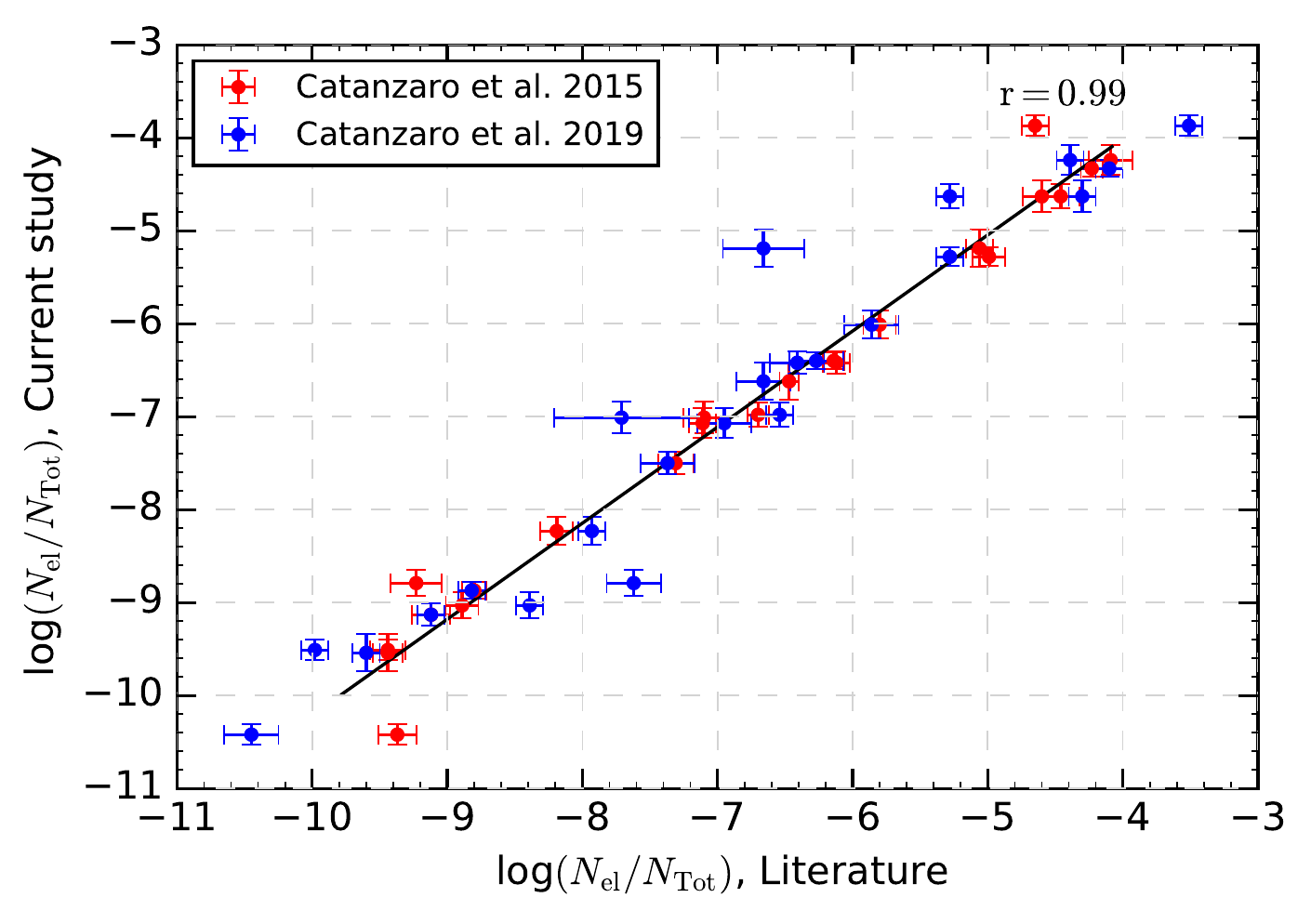}
 \caption{A comparison of our individual chemical abundances with those determined by \citet{catanzaro2015} and \citet{2019MNRAS.484.2530C}. The corresponding correlation coefficient (r) is given in the top-right corner.}
 \label{fig:comp-2015}
\end{figure*}

Calcium and scandium are found to be underabundant relative to the solar abundances, by $\approx 0.92$\,dex and $\approx 1.53$\,dex, respectively, while the heavy elements, such as strontium, yttrium, zirconium, barium, lanthanum, cerium, and neodymium are overabundant. Our abundance analysis reveals a chemical pattern typical for Am stars. As shown in Fig.\,\ref{fig:comp-2015}, our LTE abundances are in agreement with those determined by \citet{catanzaro2015} and \citet{2019MNRAS.484.2530C} in their LTE analyses.

\subsection{NLTE abundance analysis}
\label{sect:abun_nlte}

We determined the NLTE abundances for carbon, oxygen, sodium, magnesium, silicon, potassium, calcium, scandium, titanium, zinc, strontium, zirconium, and barium. These elements are some of the most easily observed elements in A-type stars. Carbon, oxygen, magnesium, silicon, and calcium are examples of $\alpha$-elements. An accurate determination of their abundances is critical for a better understanding of the history of $\alpha$-process nucleosynthesis in the universe, the formation and evolution of a large number of galaxies, as well as the physics of stars and planetary systems. The $\alpha$-process is a type of nuclear fusion in which helium is converted into heavier elements in a star \citep{narlikar1995black}. Moreover, having the best possible abundances of calcium, scandium, strontium, zirconium, and barium is also important for the correct classification of chemically peculiar stars \citep{1974ARA&A..12..257P}.

\begin{figure}
\centering
 \includegraphics[width=\columnwidth]{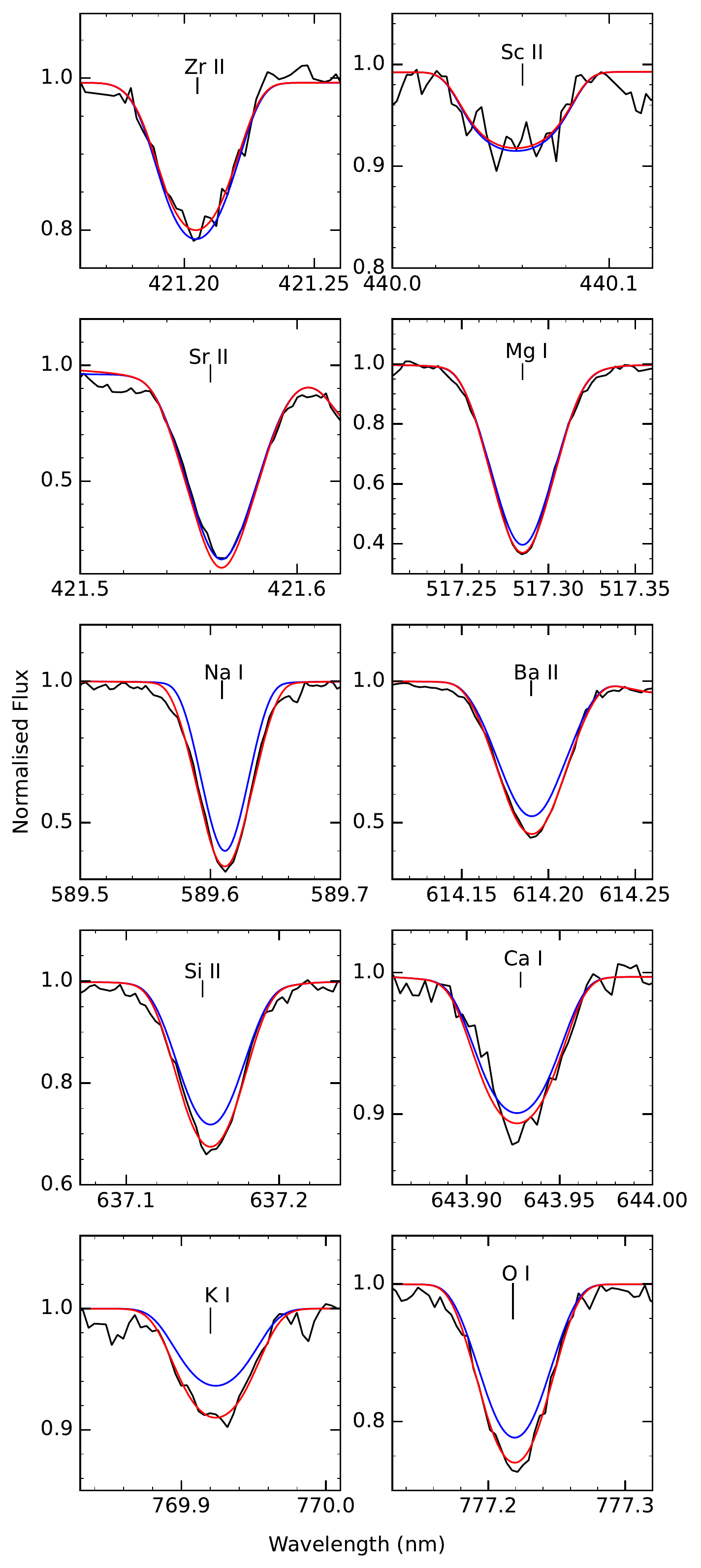}
 \caption{ NLTE ({\it red}) fits of Zr\,{\sc ii} 421.188\,nm, Sc\,{\sc ii} 440.039\,nm,  Sr\,{\sc ii} 421.552\,nm, Mg\,{\sc i} 517.285\,nm, Na\,{\sc i} 589.592\,nm, Ba\,{\sc ii} 614.171\,nm, Si\,{\sc ii} 637.137\,nm, Ca\,{\sc i} 643.929\,nm, K\,{\sc i} 769.897\,nm, and O\,{\sc i} 777.194\,nm. For each line, the LTE ({\it blue}) profile was computed with the respective abundance obtained from NLTE analysis. In each panel, the {\it black} line represents the observed spectrum. }
 \label{fig:LTE_NLTE_spectra}
\end{figure}

We used the code {\sc detail} \citep{Giddings81, Butler84} based on the accelerated $\Lambda$-iteration method \citep{rh91,rh92} to solve the radiative transfer and statistical equilibrium equations. The {\sc detail} opacity package was updated by \citet{2011JPhCS.328a2015P}. The model atoms were produced and described in detail by  \citet[][\ion{O}{i}]{sitnova_o}, \citet[][\ion{Na}{i}]{alexeeva_na}, \citet[][\ion{Mg}{i-ii}]{2018ApJ...866..153A}, \citet[][\ion{Si}{i-ii}]{2020MNRAS.493.6095M},  \citet[][\ion{K}{i}]{2020AstL...46..621N},  \citet[][\ion{Ca}{i-ii}]{mash_ca},  \citet[][\ion{Sc}{ii}]{nlte_sc2}, and \citet[][\ion{Sr}{ii}, \ion{Zr}{ii}-\ion{Zr}{iii}, \ion{Ba}{ii}]{2020MNRAS.499.3706M}.
We denote the statistical equilibrium and thermal (Saha-Boltzmann) number densities as ${\rm n_{NLTE}}$ and ${\rm n_{LTE}}$, respectively. The obtained departure coefficients (${\rm b = n_{NLTE}/n_{LTE}}$) were then used to calculate the synthetic NLTE spectrum with the code {\sc Synth}V\_NLTE \citep{2019ASPC..518..247T}. Using the visualization tool BinMag6 \citep{2018ascl.soft05015K}, we compared the synthetic NLTE and observed spectra and performed spectral line fitting. For consistency with the LTE calculations discussed in Section\,\ref{sect:abun_lte}, we used the same model atmosphere and the line list. Fig.\,\ref{fig:LTE_NLTE_spectra} shows the best NLTE fits to the selected observed lines and, for comparison, the LTE profiles computed using the abundances obtained from the NLTE analysis.

For lines of \ion{C}{i-ii}, \ion{Ti}{i-ii}, and \ion{Zn}{i}, we applied the NLTE abundance corrections predicted by \citet{2016MNRAS.462.1123A}, \citet{sitnova_ti}, and \citet{2022MNRAS.515.1510S}, respectively.

We cannot perform the NLTE calculations for the Fe-group elements V to Ni due to the absence of the model atoms, and, for the atmospheric parameter range with which we are concerned, there are no predicted NLTE abundance corrections in the literature. Having inspected the NLTE and LTE abundances published by \citet{2020MNRAS.499.3706M} for lines of \ion{Fe}{i} and \ion{Fe}{ii} in two A-type dwarf stars with \teff\ = 7250~K and 9380~K, we expect minor NLTE effects on the corresponding lines in HD~180347, of $\Delta \simeq -0.02$~dex and $\Delta <$ 0.01~dex for \ion{Fe}{i} and \ion{Fe}{ii}, respectively.

\begin{figure}
\centering
 \includegraphics[width=\columnwidth]{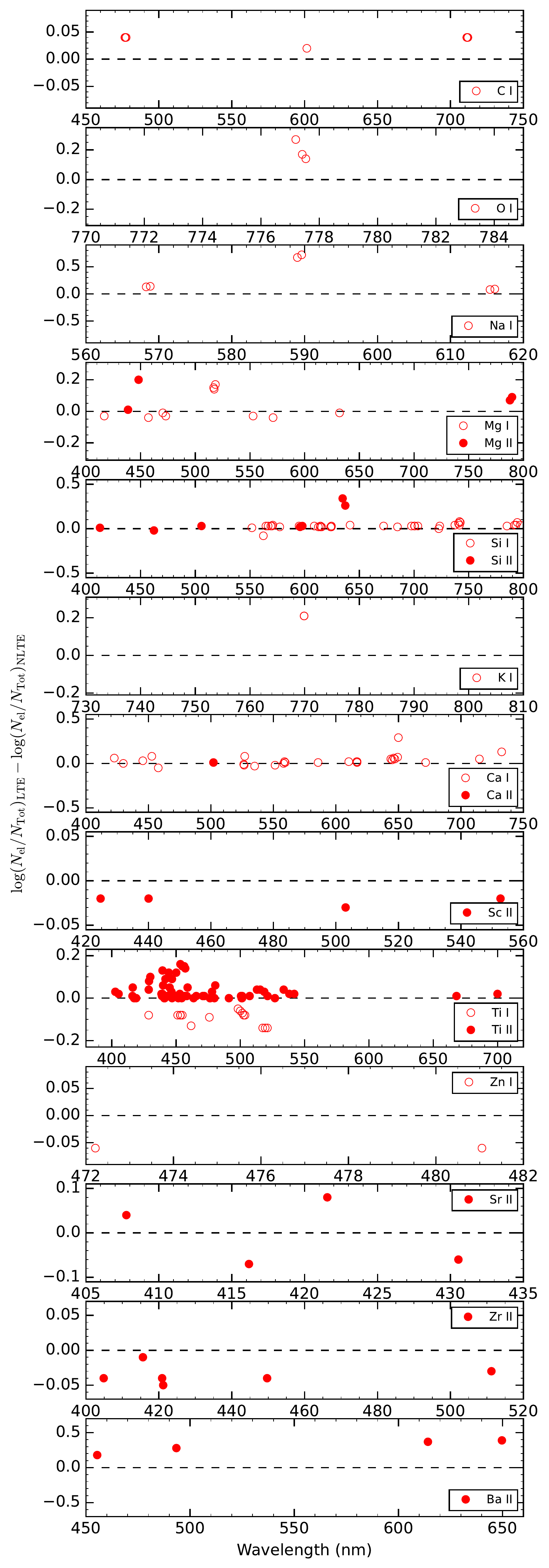}
 \caption{Abundance differences between the LTE and NLTE analyses for Na\,{\sc i},  Mg\,{\sc i}/{\sc ii}, Si\,{\sc i}/{\sc ii}, Ca\,{\sc i}/{\sc ii}, Sr\,{\sc ii}, and Ba\,{\sc ii}.}
 \label{fig:comp-LTE_NLTE}
\end{figure}

Individual line abundances are presented in Tables\,\ref{table_nlte}, \ref{table_nlte_Ca}, \ref{table_nlte_si}, and\,\ref{table_nlte_c}. Figure\,\ref{fig:comp-LTE_NLTE} shows the differences between the LTE and NLTE abundances ($\Delta$) as a function of wavelength. The NLTE effects can be different for the lines produced by the same chemical species. In general, the LTE assumption is valid in deep atmospheric layers where the medium is opaque to the continuum radiation, collisional processes in each atom prevail over radiative ones and the radiation field is close to the thermodynamic equilibrium. The departures from LTE grow towards the surface, resulting in greater NLTE effects for the strong spectral lines compared with that for the weak lines of the same chemical species. As a rule, the resonance lines and the lines arising from the low-excitation levels are stronger than the remaining lines of the same chemical species. This explains why $\Delta$ is larger for \ion{Na}{i} 588.9~nm than for \ion{Na}{i} 568.8\,nm and for \ion{Mg}{ii} 448.1~nm than for \ion{Mg}{ii} 438.5\,nm. For \ion{Mg}{i}, the weaker lines, such as 470.3~nm and 552.8\,nm, are weakened in NLTE due to the ultra-violet overionisation, resulting in slightly negative $\Delta$. In contrast, the absorption in the strong \ion{Mg}{i} 516.7, 517.2, 518.3~nm lines is larger in NLTE than in LTE, resulting in positive $\Delta$. An explanation lies with a behaviour of the source function for these strong lines that drops relative to the Planck function in the uppermost atmospheric layers \citep[see][for more details]{2018ApJ...866..153A}. For \ion{Ba}{ii}, the NLTE effects are similarly strong for the resonance lines and the low-excitation lines.

As can be seen in Tables\,\ref{table_nlte}, \ref{table_nlte_Ca}, \ref{table_nlte_si}, and\,\ref{table_nlte_c}, NLTE reduces the line-to-line scatter and thus the error of the mean abundance for most of the chemical species. This concerns, in particular, \ion{Na}{i} and \ion{Sr}{ii}.

For the elements important for identification of the star's chemical peculiarity, NLTE supports the conclusions deduced from the LTE analysis but improves the magnitudes of the abundance deviations: O, Ca, and Sc are strongly depleted relative to their solar abundances, with [O/H]$_{\rm NLTE}$ = $-0.7$\,dex, [Ca/H]$_{\rm NLTE}$ = $-0.96$\,dex, and [Sc/H]$_{\rm NLTE}$ = $-1.51$\,dex, while Sr, Zr, and Ba are enhanced, with [X/H]\,$>$\,0.7\,dex. We note that potassium, with [K/H]$_{\rm NLTE}$ = $-0.34$\,dex, falls short of solar abundance. NLTE reduces the overabundances of Na and Si obtained in LTE.

To the best of our knowledge, the NLTE abundance analyses are available in the literature for three Am stars. For a nascent Am star HD\,131399A (\teff\ = 9200\,K, \logg\ = 4.37\,cm\,s$^{-2}$), \citet{2017A&A...604L...9P} determined the NLTE abundances of He, C, N, O, Mg, Si, Ti, and Fe. In contrast to our target, HD~131399A does not reveal any underabundances of C, O, Ca, and Sc, probably, due to its young age (16\,Myr). Although Ca and Sc were treated under the LTE assumption, the NLTE effects are not expected to reduce their abundances to the level observed in HD~180347. Common for HD\,131399A and our target is a substantial enhancement in Sr, Zr, and Ba. For \teff /\logg\ of HD\,131399A, NLTE is expected to increase abundances of these elements to an even higher level. For a benchmark Am star Sirius (HD~48915, \teff\,=\,9850\,K, \logg\,=\,4.30\,cm\,s$^{-2}$) and an Am star HD~72660 with close atmospheric parameters (\teff\,=\,9700\,K, \logg\,=\,4.10\,cm\,s$^{-2}$), \citet{2020MNRAS.499.3706M} derived the NLTE abundances of He, C, O, Na, Mg, Si, Ca, Ti, Fe, Sr, Zr, Ba, and Nd. \citet{2020MNRAS.499.3706M} note a rather different behaviour of Ca and Sc in their two stars, namely, Ca is slightly depleted and Sc is strongly depleted in Sirius, while Ca is enhanced and Sc is moderately depleted in HD\,72660. Despite Sc being treated in LTE, the NLTE effects are expected to be similar for these two stars with similar atmospheric parameters. Both stars reveal enhancements in the heavy elements (Sr to Nd) at a level similar to that for our target or the higher level.

\begin{table}
\centering
\caption{Listed are atomic data for O\,{\sc i}, Na\,{\sc i}, Mg\,{\sc i/ii}, K\,{\sc i}, Sr\,{\sc i}, and Ba\,{\sc ii} lines: the wavelength in nm of the lines used in the analysis, the oscillator strengths ($\log(gf)$) and excitation energy ($E_{\rm low}$) of the lower level as given in the VALD3 database, the LTE and NLTE abundances, and the abundance correction ($\Delta$). $\Delta$ is the difference between LTE and NLTE abundances.  LTE means $\log(N_{\rm el}/N_{\rm Tot})_{\rm LTE}$ and NLTE means $\log(N_{\rm el}/N_{\rm Tot})_{\rm NLTE}$.}
\label{table_nlte}
 \begin{tabular}{lrlrrr}
\hline
\hline
\noalign{\smallskip}
Wavelength  &$\log(gf)$&$E_{\rm low}$&LTE&NLTE&$\Delta$\\
  (nm) &&(eV)&&&\\
  \hline
  O\,{\sc i}&&&&& \\
  777.194 & 0.369 & 9.15 & -3.78 & -4.05 & 0.27\\
  777.417 & 0.223 & 9.15 & -3.87 & -4.04 & 0.17\\
  777.539 & 0.002 & 9.15 & -3.95 & -4.07 & 0.14\\
\hline
Mean    &       & & -3.87 & -4.05 &  0.19\\
  $\sigma$&&&0.07&0.01&\\
 \hline
  Na\,{\sc i}&&&&& \\
  568.263 & -0.706 & 2.10 & -5.13 & -5.26 & 0.13\\
  568.82 & -0.452 & 2.10 & -5.33 & -5.47 & 0.14\\
  588.995 & 0.11 & 0.0 & -4.94 & -5.61 & 0.67\\
  589.592 & -0.194 & 0.0 & -4.93 & -5.65 & 0.72\\
  615.423 & -1.547 & 2.10 & -5.36 & -5.44 & 0.08\\
  616.075 & -1.246 & 2.10 & -5.44 & -5.53 & 0.09\\
\hline
  Mean    &       & & -5.19 & -5.49 &  0.31\\
  $\sigma$&&&0.20&0.13&\\
  \hline
  Mg\,{\sc i}&&&&&\\
  416.727 & -0.745 & 4.35 & -4.83 & -4.8 & -0.03\\
  457.11 & -5.623 & 0.0 & -4.66 & -4.62 & -0.04\\
  470.299 & -0.44 & 4.35 & -4.91 & -4.9 & -0.01\\
  473.003 & -2.347 & 4.35 & -4.27 & -4.24 & -0.03\\
  516.732 & -0.931 & 2.71 & -4.56 & -4.71 & 0.15\\
  517.268 & -0.45 & 2.71 & -4.69 & -4.83 & 0.14\\
  518.36 & -0.239 & 2.72 & -4.37 & -4.54 & 0.17\\
  552.841 & -0.498 & 4.35 & -4.89 & -4.86 & -0.03\\
  571.109 & -1.724 & 4.35 & -4.69 & -4.65 & -0.04\\
  631.875 & -2.103 & 5.11 & -4.34 & -4.33 & -0.01\\
  \hline
  Mean    &        && -4.62 & -4.65 &  0.03\\
  $\sigma$&        &&  0.22 &  0.21 &      \\
  \hline
  Mg\,{\sc ii}&&&&&\\
  438.464 & -0.79 & 10.00 & -4.6 & -4.61 & 0.01\\
  448.115 & 1.385 & 8.86 & -4.56 & -4.76 & 0.2\\
  787.705 & 0.39 & 10.00 & -4.74 & -4.81 & 0.07\\
  789.637 & 0.69 & 10.00 & -4.75 & -4.84 & 0.09\\
  \hline
  Mean    &       & & -4.66 & -4.76 &  0.09\\
  $\sigma$&        &&  0.08 &  0.09 &      \\
  \hline
  K\,{\sc i}&&&&&\\
  769.897 & -0.180 & 0.0 & -7.14 & -7.35 & -0.21\\
  \hline
  Sr\,{\sc ii}&&&&\\
  407.771 & 0.143 & 0.0 & -7.94 & -7.98 & 0.04\\
  416.179 & -0.327 & 2.94 & -8.43 & -8.36 & -0.07\\
  421.552 & -0.173 & 0.0 & -8.02 & -8.1 & 0.08\\
  430.544 & -0.041 & 3.04 & -8.52 & -8.46 & -0.06\\
\hline
  Mean    &       & & -8.23 & -8.23 &  0.00\\
  $\sigma$&       & &  0.25 &  0.19 &      \\
  \hline
  Ba\,{\sc ii}&&&&\\
  455.403 & 0.17 & 0.0 & -8.54 & -8.71 & 0.17\\
  493.408 & -0.17 & 0.0 & -8.5 & -8.76 & 0.26\\
  614.171 & -0.07 & 0.7 & -8.56 & -8.9 & 0.34\\
  649.69 & -0.37 & 0.6 & -8.37 & -8.78 & 0.39\\
\hline
  Mean    &     &   & -8.49 & -8.79 &  0.29\\
  $\sigma$&      &  &  0.07 &  0.07 &      \\
  \hline

  \end{tabular}
\end{table}

\begin{table}
\centering
\caption{Similar to Table\,\ref{table_nlte} but for Ca\,{\sc i/ii}, Sc\,{\sc ii}, and Zr\,{\sc ii} lines.}
\label{table_nlte_Ca}
 \begin{tabular}{lrlrrr}
\hline
\hline
\noalign{\smallskip}
Wavelength  &$\log(gf)$&$E_{\rm low}$&LTE&NLTE&$\Delta$\\
  (nm) &&(eV)&&&\\
  \hline
  Ca\,{\sc i}&&&&&\\
   422.673 & 0.244 & 0.0 & -6.85 & -6.9 & 0.06\\
  429.899 & -0.359 & 1.89 & -6.83 & -6.83 & 0.0\\
  445.478 & 0.258 & 1.9 & -6.53 & -6.56 & 0.03\\
  452.693 & -0.548 & 2.71 & -6.71 & -6.79 & 0.08\\
  457.855 & -0.697 & 2.52 & -6.46 & -6.41 & -0.05\\
  526.224 & -0.471 & 2.52 & -7.08 & -7.07 & -0.01\\
  526.556 & -0.113 & 2.52 & -6.36 & -6.34 & -0.02\\
  527.027 & 0.162 & 2.53 & -6.32 & -6.4 & 0.08\\
  534.947 & -0.31 & 2.71 & -6.63 & -6.6 & -0.03\\
  551.298 & -0.464 & 2.93 & -6.47 & -6.45 & -0.02\\
  558.197 & -0.555 & 2.52 & -6.48 & -6.48 & 0.0\\
  558.875 & 0.358 & 2.53 & -6.8 & -6.81 & 0.02\\
  559.446 & 0.097 & 2.52 & -6.46 & -6.48 & 0.01\\
  585.745 & 0.24 & 2.93 & -6.26 & -6.27 & 0.01\\
  610.272 & -0.793 & 1.88 & -6.5 & -6.52 & 0.02\\
  616.644 & -1.142 & 2.52 & -6.47 & -6.48 & 0.02\\
  616.905 & -0.797 & 2.52 & -6.45 & -6.46 & 0.01\\
  616.956 & -0.478 & 2.53 & -6.46 & -6.48 & 0.02\\
  643.908 & 0.39 & 2.53 & -6.74 & -6.78 & 0.05\\
  644.981 & -0.502 & 2.52 & -6.73 & -6.77 & 0.04\\
  646.257 & 0.262 & 2.52 & -6.57 & -6.63 & 0.06\\
  647.166 & -0.686 & 2.53 & -6.56 & -6.61 & 0.05\\
  649.378 & -0.109 & 2.52 & -6.75 & -6.82 & 0.07\\
  649.965 & -0.818 & 2.52 & -7.13 & -7.42 & 0.29\\
  671.768 & -0.524 & 2.71 & -6.49 & -6.5 & 0.01\\
  714.815 & 0.137 & 2.71 & -6.62 & -6.67 & 0.05\\
  732.615 & -0.208 & 2.93 & -6.7 & -6.83 & 0.13\\
 \hline
  Mean    &     &   & -6.61 & -6.64 &  0.04\\
  $\sigma$&    &    &  0.21 &  0.24 &      \\
  \hline
  Ca\,{\sc ii}&&&&&\\
  501.997 & -0.247 & 7.51 & -6.99 & -7.0 & 0.01\\
  \hline
  Sc\,{\sc ii}&&&&&\\
  424.682 & 0.24 & 0.32 & -10.48 & -10.46 & -0.02\\
  440.039 & -0.54 & 0.61 & -10.53 & -10.51 & -0.02\\
  503.102 & -0.41 & 1.36 & -10.26 & -10.23 & -0.03\\
  552.679 & -0.01 & 1.77 & -10.42 & -10.4 & -0.02\\
  \hline
  Mean    &    &    & -10.42 & -10.40 &  -0.02\\
  $\sigma$&    &    &  0.10 &  0.11 &      \\
  \hline
  Zr\,{\sc ii}&&&&&\\
  404.867 & -0.53 & 0.80 & -8.8 & -8.76 & -0.04\\
  415.628 & -0.776 & 0.71 & -8.81 & -8.8 & -0.01\\
  420.898 & -0.51 & 0.71 & -9.14 & -9.1 & -0.04\\
  421.188 & -1.04 & 0.53 & -8.86 & -8.81 & -0.05\\
  449.696 & -0.89 & 0.71 & -8.86 & -8.82 & -0.04\\
  511.227 & -0.85 & 1.66 & -8.75 & -8.72 & -0.03\\
  \hline
  Mean    &    &    & -8.87 & -8.84 &  -0.04\\
  $\sigma$&    &    &  0.13 &  0.12 &      \\
  \hline
\end{tabular}
\end{table}

\begin{table}
\centering
\caption{ Similar to Table\,\ref{table_nlte} but for Si\,{\sc i/ii} lines.}
\label{table_nlte_si}
 \begin{tabular}{lclrrr}
\hline
\hline
\noalign{\smallskip}
Wavelength  &$\log(gf)$&$E_{\rm low}$&LTE&NLTE&$\Delta$\\
  (nm) &&(eV)&&&\\
  \hline
  Si\,{\sc i}&&&&&\\
 551.753 & -2.61 & 5.08 & -4.67 & -4.68 & 0.01\\
  562.222 & -2.606 & 4.93 & -4.52 & -4.44 & -0.08\\
  564.561 & -2.14 & 4.93 & -4.27 & -4.3 & 0.03\\
  566.555 & -2.04 & 4.92 & -4.15 & -4.18 & 0.03\\
  569.043 & -1.87 & 4.93 & -4.22 & -4.25 & 0.03\\
  570.11 & -2.05 & 4.93 & -4.04 & -4.07 & 0.03\\
  570.84 & -1.47 & 4.95 & -4.15 & -4.19 & 0.04\\
  577.215 & -1.75 & 5.08 & -4.17 & -4.19 & 0.02\\
  594.854 & -1.23 & 5.08 & -4.25 & -4.28 & 0.03\\
  608.781 & -1.815 & 5.87 & -3.99 & -4.02 & 0.03\\
  612.502 & -1.465 & 5.61 & -4.13 & -4.15 & 0.02\\
  614.248 & -1.296 & 5.62 & -4.41 & -4.43 & 0.02\\
  614.502 & -1.311 & 5.62 & -4.29 & -4.32 & 0.03\\
  615.513 & -0.755 & 5.62 & -4.44 & -4.46 & 0.02\\
  623.732 & -0.975 & 5.61 & -4.42 & -4.44 & 0.02\\
  624.382 & -1.244 & 5.62 & -4.32 & -4.34 & 0.02\\
  624.447 & -1.091 & 5.62 & -4.4 & -4.43 & 0.03\\
  641.498 & -1.036 & 5.87 & -4.23 & -4.27 & 0.04\\
  672.185 & -1.527 & 5.86 & -3.9 & -3.93 & 0.03\\
  684.858 & -1.528 & 5.86 & -4.17 & -4.19 & 0.02\\
  697.651 & -1.17 & 5.95 & -4.05 & -4.08 & 0.03\\
  700.357 & -0.89 & 5.96 & -4.15 & -4.18 & 0.03\\
  700.588 & -0.69 & 5.98 & -4.25 & -4.28 & 0.03\\
  703.49 & -0.88 & 5.87 & -4.19 & -4.22 & 0.03\\
  722.621 & -1.51 & 5.61 & -4.25 & -4.25 & 0.0\\
  723.533 & -1.49 & 5.62 & -4.0 & -4.03 & 0.03\\
  737.3 & -1.18 & 5.98 & -4.13 & -4.17 & 0.04\\
  740.577 & -0.82 & 5.61 & -4.19 & -4.25 & 0.06\\
  741.536 & -1.76 & 5.62 & -4.14 & -4.18 & 0.04\\
  741.596 & -2.777 & 3.71 & -4.15 & -4.23 & 0.08\\
  742.35 & -0.176 & 5.62 & -4.75 & -4.82 & 0.07\\
  784.997 & -0.714 & 6.19 & -4.18 & -4.21 & 0.03\\
  791.838 & -0.61 & 5.95 & -4.29 & -4.33 & 0.04\\
  793.235 & -0.47 & 5.96 & -4.21 & -4.25 & 0.04\\
  794.4 & -0.31 & 5.98 & -4.2 & -4.27 & 0.07\\
  797.031 & -1.47 & 5.96 & -3.95 & -3.99 & 0.04\\
  \hline
  Mean    &    &    & -4.23 & -4.26 &  -0.03\\
  $\sigma$&    &    &  0.18 &  0.17 &      \\
  \hline
  Si\,{\sc ii}&&&&&\\
  412.805 & 0.41 & 9.84 & -4.68 & -4.69 & 0.01\\
  462.172 & -0.38 & 12.53 & -4.23 & -4.21 & -0.02\\
  505.632 & 0.53 & 10.07 & -4.34 & -4.37 & 0.03\\
  595.756 & -0.26 & 10.07 & -4.12 & -4.14 & 0.02\\
  597.893 & 0.04 & 10.07 & -4.37 & -4.4 & 0.03\\
  634.711 & 0.17 & 8.12 & -4.18 & -4.52 & 0.34\\
  637.137 & -0.04 & 8.12 & -4.29 & -4.55 & 0.26\\
\hline
  Mean    &    &    & -4.32 & -4.41 &  0.1\\
  $\sigma$&    &    &  0.17 &  0.18 &      \\
  \hline
\end{tabular}
\end{table}

\begin{table}
\centering
\caption{Similar to Table\,\ref{table_nlte} but for C\,{\sc i}, Ti\,{\sc i/ii}, and Zn\,{\sc i} lines. The NLTE abundances are obtained by adding the NLTE abundance corrections
predicted by \citet{2016MNRAS.462.1123A}, \citet{sitnova_ti}, and \citet{2022MNRAS.515.1510S} for lines \ion{C}{i}, \ion{Ti}{i-ii}, and \ion{Zn}{i}, respectively. }
\label{table_nlte_c}
 \begin{tabular}{lrlrrr}
\hline
\hline
\noalign{\smallskip}
Wavelength  &$\log(gf)$&$E_{\rm low}$&LTE&NLTE&$\Delta$\\
  (nm) &&(eV)&&&\\
  \hline
  C\,{\sc i}&&&&& \\
  476.666 & -2.617 & 7.48 & -3.88 & -3.92 & 0.04\\
  477.002 & -2.437 & 7.48 & -4.4 & -4.44 & 0.04\\
  477.589 & -2.304 & 7.49 & -4.25 & -4.29 & 0.04\\
  601.483 & -1.584 & 8.64 & -4.19 & -4.21 & 0.02\\
  711.146 & -1.09 & 8.64 & -4.3 & -4.34 & 0.04\\
  711.517 & -0.93 & 8.64 & -4.47 & -4.51 & 0.04\\
  711.965 & -1.148 & 8.64 & -4.32 & -4.36 & 0.04\\
  \hline
  Mean    &       & & -4.26 & -4.30 &  0.04\\
  $\sigma$&&&0.18&0.18&\\
  \hline
Ti\,{\sc i}&&&&& \\
  428.74 & -0.37 & 0.84 & -7.67 & -7.59 & -0.08\\
  451.273 & -0.4 & 0.84 & -6.94 & -6.86 & -0.08\\
  453.324 & 0.54 & 0.85 & -7.0 & -6.92 & -0.08\\
  453.478 & 0.35 & 0.84 & -7.48 & -7.4 & -0.08\\
  454.876 & -0.28 & 0.83 & -7.15 & -7.07 & -0.08\\
  461.727 & 0.44 & 1.75 & -7.2 & -7.07 & -0.13\\
  475.927 & 0.59 & 2.26 & -7.3 & -7.21 & -0.09\\
  498.173 & 0.57 & 0.85 & -7.27 & -7.22 & -0.05\\
  499.95 & 0.32 & 0.83 & -6.87 & -6.81 & -0.06\\
  501.616 & -0.48 & 0.85 & -7.2 & -7.13 & -0.07\\
  502.557 & 0.25 & 2.04 & -6.88 & -6.8 & -0.08\\
  503.646 & 0.14 & 1.44 & -7.09 & -7.01 & -0.08\\
  517.374 & -1.06 & 0.0 & -6.93 & -6.79 & -0.14\\
  519.297 & -0.95 & 0.02 & -6.71 & -6.57 & -0.14\\
  521.038 & -0.82 & 0.05 & -6.99 & -6.85 & -0.14\\
\hline
Mean    &       & & -7.11 & -7.02 &  -0.09\\
  $\sigma$&&&0.24&0.25&\\
  \hline
   Ti\,{\sc ii}&&&&& \\
  402.834 & -0.92 & 1.89 & -7.23 & -7.26 & 0.03\\
  405.382 & -1.07 & 1.89 & -7.38 & -7.4 & 0.02\\
  416.153 & -2.09 & 1.08 & -7.08 & -7.09 & 0.01\\
  416.364 & -0.13 & 2.59 & -7.12 & -7.17 & 0.05\\
  417.407 & -1.26 & 2.6 & -7.29 & -7.29 & 0.0\\
  419.023 & -3.122 & 1.08 & -6.96 & -6.96 & 0.0\\
  428.787 & -1.79 & 1.08 & -6.98 & -7.02 & 0.04\\
  429.022 & -0.87 & 1.16 & -7.1 & -7.18 & 0.08\\
  430.005 & -0.46 & 1.18 & -6.82 & -6.92 & 0.1\\
  438.684 & -0.96 & 2.6 & -7.08 & -7.1 & 0.02\\
  439.102 & -2.3 & 1.23 & -6.72 & -6.73 & 0.01\\
  439.406 & -1.77 & 1.22 & -7.34 & -7.36 & 0.02\\
  439.503 & -0.54 & 1.08 & -7.18 & -7.31 & 0.13\\
  439.584 & -1.93 & 1.24 & -7.28 & -7.29 & 0.01\\
  439.977 & -1.2 & 1.24 & -7.2 & -7.26 & 0.06\\
  440.924 & -2.78 & 1.24 & -6.82 & -6.82 & 0.0\\
  440.952 & -2.53 & 1.23 & -6.89 & -6.89 & 0.0\\
  441.107 & -0.65 & 3.09 & -7.1 & -7.11 & 0.01\\
  441.193 & -2.62 & 1.22 & -6.95 & -6.95 & 0.0\\
  441.771 & -1.19 & 1.16 & -7.3 & -7.39 & 0.09\\
  441.833 & -1.99 & 1.24 & -7.21 & -7.22 & 0.01\\
  442.194 & -1.64 & 2.06 & -7.11 & -7.12 & 0.01\\
  444.173 & -2.33 & 1.18 & -7.17 & -7.18 & 0.01\\
  444.38 & -0.71 & 1.08 & -7.21 & -7.33 & 0.12\\
  445.048 & -1.52 & 1.08 & -6.87 & -6.92 & 0.05\\
  446.445 & -1.81 & 1.16 & -7.19 & -7.22 & 0.03\\
  446.85 & -0.63 & 1.13 & -7.11 & -7.2 & 0.09\\
  446.915 & -2.55 & 1.08 & -6.91 & -6.91 & 0.0\\
  447.085 & -2.02 & 1.16 & -7.21 & -7.23 & 0.02\\
  448.832 & -0.5 & 3.12 & -7.18 & -7.19 & 0.01\\
  450.127 & -0.77 & 1.12 & -7.11 & -7.23 & 0.12\\
  \hline
\end{tabular}
\end{table}

\begin{table}
\centering
\contcaption{.}
\label{table_nlte_c2}
 \begin{tabular}{lrlrrr}
\hline
\hline
\noalign{\smallskip}
Wavelength  &$\log(gf)$&$E_{\rm low}$&LTE&NLTE&$\Delta$\\
  (nm) &&(eV)&&&\\
 \hline
Ti\,{\sc ii}&&&&& \\
  451.833 & -2.56 & 1.08 & -7.12 & -7.12 & 0.0\\
  452.947 & -1.75 & 1.57 & -7.1 & -7.12 & 0.02\\
  453.396 & -0.53 & 1.24 & -7.13 & -7.29 & 0.16\\
  454.402 & -2.58 & 1.24 & -7.15 & -7.15 & 0.0\\
  454.962 & -0.22 & 1.58 & -6.89 & -7.04 & 0.15\\
  456.376 & -0.795 & 1.22 & -7.27 & -7.42 & 0.15\\
  456.831 & -3.03 & 1.22 & -7.02 & -7.03 & 0.01\\
  457.197 & -0.31 & 1.57 & -6.8 & -6.94 & 0.14\\
  458.341 & -2.84 & 1.16 & -7.01 & -7.02 & 0.01\\
  458.996 & -1.62 & 1.24 & -7.25 & -7.3 & 0.05\\
  463.632 & -3.024 & 1.24 & -7.15 & -7.15 & 0.0\\
  465.72 & -2.29 & 1.24 & -6.92 & -6.93 & 0.01\\
  470.866 & -2.35 & 1.24 & -7.23 & -7.24 & 0.01\\
  471.952 & -3.32 & 1.24 & -6.93 & -6.94 & 0.01\\
  476.388 & -2.4 & 1.22 & -6.8 & -6.8 & 0.0\\
  476.453 & -2.69 & 1.24 & -6.96 & -6.96 & 0.0\\
  477.999 & -1.26 & 3.28 & -7.28 & -7.31 & 0.03\\
  479.853 & -2.66 & 1.08 & -7.29 & -7.29 & 0.0\\
  480.509 & -0.96 & 2.06 & -7.23 & -7.29 & 0.06\\
  491.119 & -0.64 & 3.12 & -7.26 & -7.26 & 0.0\\
  500.516 & -2.73 & 1.57 & -7.07 & -7.08 & 0.01\\
  501.021 & -1.35 & 3.09 & -7.06 & -7.06 & 0.0\\
  501.333 & -2.028 & 3.09 & -6.91 & -6.91 & 0.0\\
  501.369 & -2.14 & 1.58 & -7.18 & -7.19 & 0.01\\
  507.229 & -1.02 & 3.12 & -6.89 & -6.9 & 0.01\\
  512.916 & -1.34 & 1.89 & -6.77 & -6.81 & 0.04\\
  515.407 & -1.75 & 1.57 & -7.19 & -7.23 & 0.04\\
  518.591 & -1.41 & 1.89 & -7.27 & -7.3 & 0.03\\
  521.154 & -1.41 & 2.59 & -7.2 & -7.21 & 0.01\\
  526.861 & -1.61 & 2.6 & -7.04 & -7.04 & 0.0\\
  533.679 & -1.6 & 1.58 & -7.21 & -7.25 & 0.04\\
  538.102 & -1.97 & 1.57 & -7.1 & -7.12 & 0.02\\
  541.877 & -2.13 & 1.58 & -7.17 & -7.19 & 0.02\\
  668.013 & -1.89 & 3.09 & -7.02 & -7.03 & 0.01\\
  699.891 & -1.28 & 3.12 & -6.99 & -7.01 & 0.02\\
\hline
Mean    &       & & -7.09 & -7.12 &  0.03\\
  $\sigma$&&&0.16&0.16&\\
 \hline
 Zn\,{\sc i}&&&&& \\
4722.153  &   -0.338	&4.03 & -6.92  &-6.86 & -0.06\\
4810.528   &  -0.137	&4.08 & -7.04  &-6.98 & -0.06\\
\hline
  Mean    &       & & -6.98 & -6.92 &  -0.06\\
  $\sigma$&&&0.06&0.06&\\
\hline
\end{tabular}
\end{table}

\section{Conclusion}
\label{concl}

\citet{2019MNRAS.484.2530C} suspected that HD\,180347 had $\delta$ Scuti-type pulsations. Based on the diffusion theory, such pulsations are unexpected given that it is a candidate Am star. Therefore, the photometric variability and chemical nature of HD\,180347 are investigated in detail in this study. The analysis is based on high-precision photometric data by \tess\ and a high-resolution spectrum obtained with HERMES mounted at the Cassegrain focus of the 1.2-m Mercator telescope located at La Palma, Spain. This study increases the statistics of Am stars with accurate NLTE abundances derived for important chemical elements. The knowledge of the abundance anomalies depending on stellar parameters should play a key role in understanding the mechanisms of the chemical peculiarity of Am stars.

We searched for the variability type, determined the spectral type, and obtained fundamental parameters such as the effective temperature, surface gravity, and projected rotational, microturbulent, and radial velocity. Based on these results, we estimated the radius, equatorial rotational velocity, inclination angle, mass, and age of \target. Finally, we performed a detailed chemical abundance analysis.

Based on the variability analysis of the
high-precision photometric data by \tess\ (sectors 14, 15, and 26), we classify \target\ as a rotational variable with a period of 4.1\,$\pm$\,0.2\,days. In reference to the observation limit of \tess, no pulsations were detected as previously suspected by \citet{2019MNRAS.484.2530C}. The rotational amplitude of 0.08\,mmag indicates significant spots and thus potentially significant magnetic fields. A spectropolarimetric analysis of this star is required to estimate its magnetic field strength.

We calculated LTE abundances for 25 different chemical elements. We also report the non-local thermodynamic equilibrium (NLTE) abundances for 13 of them, including Ca, Sc, Sr, Zr, and Ba, which are significant for the characterization of chemical peculiarity. The use of NLTE increases the accuracy of the derived abundances and indicates that Ca and Sc are depleted in HD\,180347 relative to their solar abundances, while heavy elements beyond Sr are increased by more than 0.7\,dex. Based on the spectral classification analysis and chemical abundance pattern, we classified this star as Am (kA1hA8mA8).

The availability of HERMES and other spectrographs capable of high-resolution spectroscopy will allow us, in future, to continue with the study of A and probable Am stars especially those reported in \citet{2009A&A...498..961R}. Efforts will be made to increase the number of elements in NLTE abundance analyses.

\section*{Acknowledgments}

The International Science Program (ISP) of Uppsala University and the African Astronomical Society (AfAS) financed the study. The part of the work presented here is supported by the Belgo-Indian Network for Astronomy \& Astrophysics (BINA), approved by the International Division, Department of Science and Technology (DST, Govt. of India; DST/INT/Belg/P09/2017) and the Belgian Federal Science Policy Office (BELSPO, Govt. of Belgium; BL/33/IN12). SJ acknowledges the financial support received from the BRICS grant DST/IC/BRICS/2023/5. We thank Dr Yves Fr\'{e}mat for kindly providing the \textsc{girfit} code. The authors acknowledge the anonymous reviewer for his/her careful reading of our manuscript and  for many insightful comments and suggestions which improved the paper. Based on observations obtained with the HERMES spectrograph, which is supported by the Research Foundation - Flanders (FWO), Belgium, the Research Council of KU Leuven, Belgium, the Fonds National de la Recherche Scientifique (F.R.S.-FNRS), Belgium, the Royal Observatory of Belgium, the Observatoire de Genève, Switzerland and the Thüringer Landessternwarte Tautenburg, Germany. This publication makes use of VOSA, developed under the Spanish Virtual Observatory project supported by the Spanish MINECO through grant AyA2017-84089. VOSA has been partially updated by using funding from the European Union's Horizon 2020 Research and Innovation Programme, under Grant Agreement 776403 (EXOPLANETS-A). This work has made use of the VALD database, operated at Uppsala University, the Institute of Astronomy RAS in Moscow, and the University of Vienna.

\section*{DATA AVAILABILITY}
The data underlying this article will be shared on reasonable request to the corresponding author.

 \bibliographystyle{mnras}
 \bibliography{ref}

\bsp    
\label{lastpage}
\end{document}